\documentclass{article}
\usepackage{iclr2023_conference_costum_preprint,times}

\usepackage{amsmath,amsfonts,bm}

\def\eqref#1{equation~\ref{#1}}

\def\1{\bm{1}}

\DeclareMathAlphabet{\mathsfit}{\encodingdefault}{\sfdefault}{m}{sl}
\SetMathAlphabet{\mathsfit}{bold}{\encodingdefault}{\sfdefault}{bx}{n}

\usepackage[utf8]{inputenc} 
\usepackage[T1]{fontenc}    
\usepackage{hyperref}      
\usepackage{url}           
\usepackage{booktabs}
\usepackage{amsfonts,amsmath,amssymb}
\usepackage{bm} 
\usepackage{bbm} 
\usepackage{mathtools}

\usepackage[para]{threeparttable}

\usepackage{tocloft}
\newcommand{\stoptocwriting}{
  \addtocontents{toc}{\protect\setcounter{tocdepth}{-5}}}
\newcommand{\resumetocwriting}{
  \addtocontents{toc}{\protect\setcounter{tocdepth}{\arabic{tocdepth}}}}

\usepackage{nicefrac}       
\usepackage{microtype}  
\usepackage{xcolor} 
\usepackage{ml-defs}
\usepackage{float}

\usepackage{subfig}

\definecolor{codeblue}{RGB}{64,165,155}
\definecolor{mathblue}{RGB}{71,184,172}
\definecolor{JKUblue}{RGB}{0,132,187}
\definecolor{JKUgreen}{RGB}{91,167,85} 
\definecolor{JKUpurple}{RGB}{174,97,157} 
\hypersetup{
   linkcolor=JKUblue,
   citecolor=JKUblue,
   filecolor=JKUblue,
   urlcolor=JKUpurple,
   colorlinks=true
}

\title{Context-enriched molecule representations improve few-shot drug discovery}

\author{Johannes Schimunek$^{1}$, Philipp Seidl$^{1}$, Lukas Friedrich$^{2}$, Daniel Kuhn$^{2}$,\\ \textbf{Friedrich Rippmann}$^{2}$\textbf{, Sepp Hochreiter}$^{1}$\textbf{, and Günter Klambauer}$^{1}$\\
$^{1}$ ELLIS Unit Linz and LIT AI Lab, Institute for Machine Learning, Johannes Kepler University\\
\phantom{$^{1}$} Linz, Austria\\
$^{2}$ Computational Chemistry \& Biologics, Merck Healthcare, Darmstadt, Germany\\
}

\iclrfinalcopy 

\stoptocwriting
\begin{document}

\maketitle

\begin{abstract}
A central task in computational drug discovery is 
to construct models from known active molecules 
to find further promising molecules for subsequent screening. 
However, 
typically only very few active molecules are known. 
Therefore, few-shot learning methods 
have the potential to improve the effectiveness 
of this critical phase of the drug discovery process. 
We introduce a new method for few-shot drug discovery. 
Its main idea is to enrich a molecule representation by 
knowledge about known context or reference molecules. 
Our novel concept for molecule representation enrichment 
is to associate molecules from both the support set and the 
query set with a large set of reference (context) molecules
through a modern Hopfield network. 
Intuitively, this enrichment step is analogous 
to a human expert who would associate a given 
molecule with familiar molecules whose properties are known. 
The enrichment step reinforces and amplifies  
the covariance structure of the data, while simultaneously 
removing spurious correlations arising from the decoration of molecules. 
Our approach is compared
with other few-shot methods for drug discovery 
on the FS-Mol benchmark dataset.
On FS-Mol, our approach outperforms all compared methods 
and therefore sets a new state-of-the art for 
few-shot learning in drug discovery.
An ablation study shows that the enrichment step 
of our method
is the key to improve the predictive quality. 
In a domain shift experiment, 
we further demonstrate the robustness of our method.
Code is available at \url{https://github.com/ml-jku/MHNfs}.
\end{abstract}

\section{Introduction}
To improve human health, combat diseases, and tackle pandemics
there is a steady need of discovering new drugs in a fast and efficient way.
However, the drug discovery process is time-consuming and cost-intensive \citep{arrowsmith2011phase}.
Deep learning methods have
been shown to reduce time and costs of this process \citep{chen2018rise,walters2021critical}.
They diminish the required number of both wet-lab measurements and
molecules that must be synthesized  \citep{merk2018novo,schneider2020rethinking}. 
However, as of now, deep learning approaches 
use only the molecular information about 
the ligands after being trained on a large training set.
At inference time, they yield highly accurate property 
and activity prediction \citep{mayr2018large,yang2019analyzing}, 
generative \citep{segler2018generating,gomez2018automatic}, 
or synthesis models \citep{segler2018planning,seidl2022improving}.

\textbf{Deep learning methods in drug discovery
usually require large amounts of biological measurements.}
To train deep learning-based activity and property prediction models 
with high predictive performance,
hundreds or thousands of data points per task are required. 
For example, well-performing predictive models 
for activity prediction tasks of ChEMBL have been trained 
with an average of 3,621 activity points per task -- i.e., drug target -- by \citet{mayr2018large}.
The ExCAPE-DB dataset provides on average 42,501 measurements per task \citep{sun2017excape,sturm2020industry}.
\citet{wu2018moleculenet} published a large scale benchmark for molecular machine learning, 
including prediction models for the SIDER dataset \citep{kuhn2016sider}
with an average of 5,187 data points, 
Tox21 \citep{huang2016modelling,mayr2016deeptox} with on average 9,031,
and ClinTox \citep{wu2018moleculenet} with 1,491 measurements per task.
However, for typical drug design projects,
the amount of available measurements is very limited \citep{stanley2021fs, waring2015analysis,hochreiter2018machine}, since 
in-vitro experiments are expensive and time-consuming. 
Therefore, methods that need only few measurements to 
build precise prediction models are desirable.
This problem -- i.e., the challenge of learning from few data points -- is the focus 
of machine learning areas like 
meta-learning \citep{schmidhuber1987evolutionary,bengio1990learning,hochreiter2001learning} 
and few-shot learning \citep{miller2000learning,bendre2020learning, wang2020generalizing}.

\textbf{Few-shot learning tackles the low-data problem that is ubiquitous in drug discovery.} 
Few-shot learning methods have been predominantly 
developed and tested on image datasets \citep{bendre2020learning, wang2020generalizing}
and have recently been adapted to drug discovery problems \citep{altae2017low,guo2021few,wang2021property,stanley2021fs,chen2022meta}.
They are usually categorized into three groups according 
to their main approach \citep{bendre2020learning, wang2020generalizing,adler2020cross}. 
a) Data-augmentation-based approaches augment the available samples and generate new, more diverse data points \citep{chen2020simple, zhao2019data, antoniou2019assume}.  
b) Embedding-based and nearest neighbour approaches \textcolor{black}{learn embedding space representations. Predictive models 
can then be constructed from only few data points 
by comparing these embeddings.} 
For example, in Matching Networks \citep{vinyals2016matching} an attention mechanism that relies on embeddings is the basis for the predictions.
Prototypical Networks \citep{snell2017prototypical} 
create prototype representations for each class 
using the above mentioned representations in the embedding space. 
c) Optimization-based or fine-tuning methods utilize 
a meta-optimizer that focuses on 
efficiently navigating the parameter space.
For example, with MAML the meta-optimizer learns initial weights 
that can be adapted to a novel task by few optimization steps \citep{finn2017model}.

Most of these approaches have already been applied to
few-shot drug discovery (see Section~\ref{sec:relatedwork}).
Surprisingly, almost all these few-shot learning methods in drug discovery 
are worse than a naive baseline, which does not even use
the support set (see Section~\ref{sec:experiments}). 
We hypothesize 
that the under-performance of these methods 
stems from disregarding the context -- 
both in terms 
of similar molecules and similar activities.
Therefore, we propose a method that informs the representations
of the query and support set with a large number 
of context molecules covering the chemical space.  

\textbf{Enriching molecule representations with context using associative memories.} 
In data-scarce situations, 
humans extract co-occurrences and covariances
by associating current perceptions with memories \citep{Bonner:21,Potter:12}.
When we show a small set of
active molecules to a human expert in drug discovery, 
the expert associates them with 
known molecules to suggest further active molecules \citep{gomez2018decision,he2021molecular}. 
In an analogous manner, our novel concept for few-shot
learning uses associative memories to extract co-occurrences
and the covariance structure of the original data
and to amplify them in the representations \citep{furst2021cloob}. 
We use Modern Hopfield Networks (MHNs) as an associative memory, since they
can store a large set of
context molecule representations \citep[Theorem 3]{ramsauer2021hopfield}.
The representations that are retrieved from the MHNs 
replace the original representations of the query and support 
set molecules. 
Those retrieved representations have
amplified co-occurrences and covariance structures, while  
peculiarities and spurious co-occurrences of the query and support 
set molecules are averaged out. 

In this work, our contributions are the following:
\begin{itemize}\itemsep0em 
    \item We propose a new architecture \textbf{MHNfs} for few-shot learning in 
    drug discovery.
    \item We achieve a new state-of-the-art on the
    benchmarking dataset FS-Mol.
    \item We introduce a novel concept to enrich the molecule 
    representations with context by associating them with a large 
    set of context molecules.
    \item We add a naive baseline to the FS-Mol benchmark that 
    yields better results than almost all other published few-shot learning methods.
    \item We provide results of an ablation study and a domain shift experiment 
    to further demonstrate the effectiveness of our new method. 
\end{itemize}

\section{Problem setting} 
\label{sec:problem}
Drug discovery projects 
revolve around models $g(\Bm) $ that can 
predict a molecular property or activity $\hat y$,
given 
a representation $\Bm$ of 
an input molecule from a chemical 
space $\mathcal M$. We consider machine learning
models $\hat y = g_{\Bw}(\Bm)$ with 
parameters $\Bw$ that have been selected using
a training set. 
Typically, deep learning based property prediction uses 
a molecule encoder 
$f^{\mathrm{ME}}:\mathcal M \rightarrow \mathbb{R}^d$.
The molecule encoder can process different
symbolic or low-level representations of molecules, 
such as molecular descriptors \citep{bender2004similarity,unterthiner2014deep,mayr2016deeptox},
SMILES \citep{weininger1988smiles,mayr2018large,winter2019learning, segler2018generating},
or molecular graphs \citep{merkwirth2005automatic,kearnes2016molecular, yang2019analyzing, jiang2021could} 
and can be pre-trained on related property 
prediction tasks. 

For few-shot learning, the goal is to select
a high-quality predictive model based on a small set 
of molecules $\{\Bx_1,\ldots, \Bx_N\}$ 
with associated measurements $\By=\{y_1,\ldots,y_N\}$.
The measurements are usually assumed to be binary 
$y_n \in \{-1,1\}$, corresponding to the molecule being inactive or active.
The set
$\{(\Bx_n,y_n)\}_{n=1}^N$
is called the \emph{support set}
that contains samples from a prediction task and $N$ is the \emph{support 
set size}.
The goal is to construct a model that correctly predicts $y$ for an $\Bx$
that is not in the support set -- in other words, 
a model that generalizes well.
 
Standard supervised machine learning approaches \textcolor{black}{typically just show limited predictive power} at this task \citep{stanley2021fs} \textcolor{black}{since they tend to overfit on the support set due to a small number of training samples}.
These approaches
learn the parameters $\Bw$ of the model $g_{\Bw}$ 
from the support set in a supervised manner.
However, they heavily overfit to the support set when $N$ is small.
Therefore, few-shot learning methods are necessary to 
construct models from the support set that generalize well to new data.

\section{MHNfs: Hopfield-based molecular context enrichment for few-shot drug discovery}
\label{sec:method}

We aim at increasing 
the generalization capabilities of few-shot
learning methods in drug discovery by 
enriching the molecule representations with molecular 
context.
In comparison to the support set, which encodes 
information about the task, the context set -- i.e. a 
large set of molecules -- includes information about a large chemical space. The query and the support set 
molecules perform a retrieval from the context set and 
thereby enrich their representations. 
We detail this in the following.

\subsection{Model architecture}

\begin{figure}
    \centering
    \includegraphics[width=1.\textwidth]{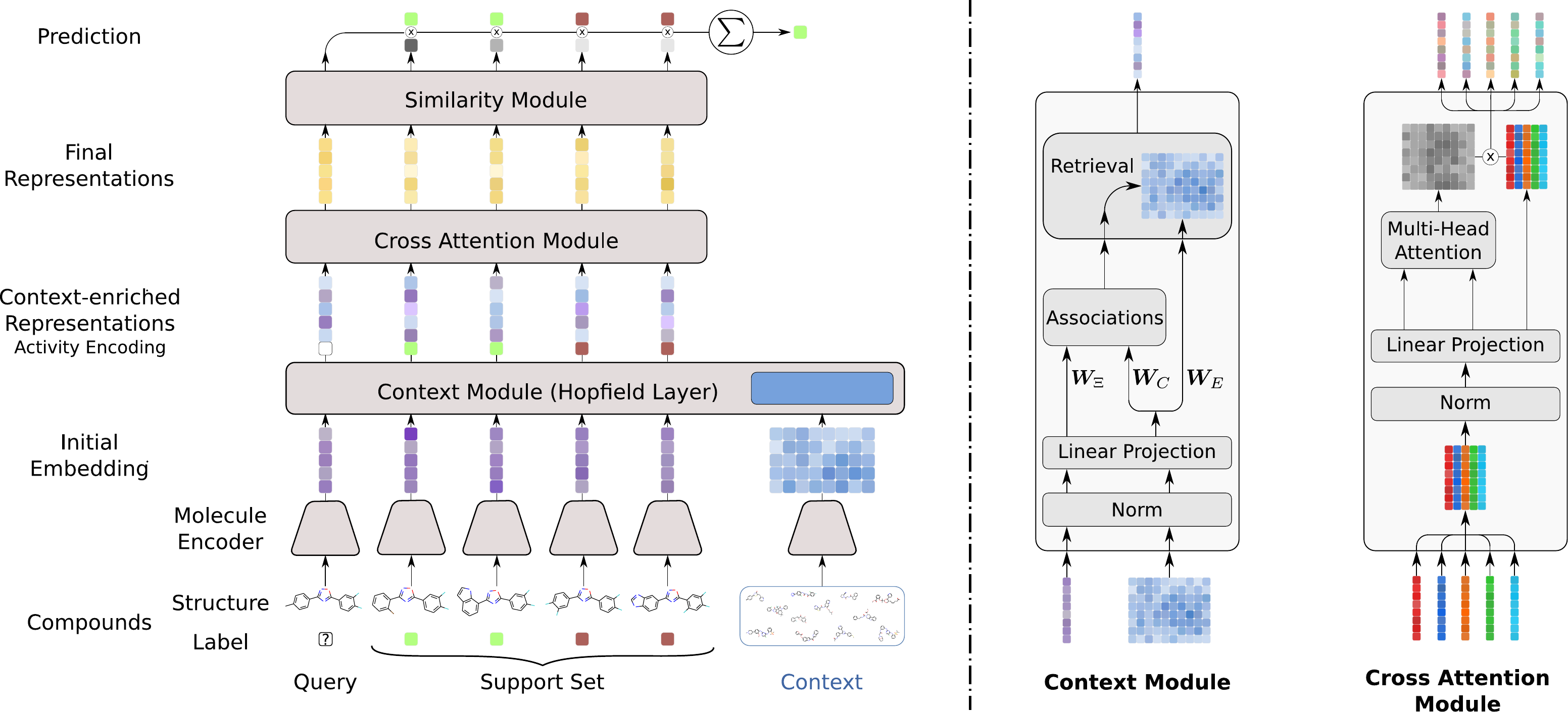}
    \caption{Schematic overview of our architecture. 
    \textbf{Left:} All molecules are fed through a shared 
    molecule encoder to obtain embeddings. Then, the context module (CM)
    enriches the representations by associating them with context molecules.
    The cross-attention module (CAM) enriches representations by mutually associating
    the query and support set molecules. Finally, the similarity module
    computes the prediction for the query molecule. \textbf{Right:} 
    Detailed depiction of the operations in the CM and
    the CAM.}
    \label{fig:overview_model}
\end{figure}

We propose an architecture which consists of three consecutive modules. 
The first module -- a) the \emph{context module} $f^{\mathrm{CM}}$ --
enriches molecule representations by retrieving from a large 
set of molecules. 
The second module -- b) the \emph{cross-attention module} $f^{\mathrm{CAM}}$ \citep{hou2019cross, chen2021sparse} -- 
enables the effective exchange of information 
between the query molecule and the support set molecules. 
Finally the prediction for the query molecule 
is computed by using the usual 
c) \emph{similarity module} $f^{\mathrm{SM}}$ \citep{koch2015siamese,altae2017low}:
\begin{align}
\text{context module:}& \hspace{-1.7cm} &  \Bm'&=f^{\mathrm{CM}}(\Bm, \BC) \nonumber \\
                      & \hspace{-1.7cm} &  \BX'&=f^{\mathrm{CM}}(\BX, \BC), \\
\text{cross-attention module:}& \hspace{-1.7cm} &  [\Bm'',\BX'']&=f^{\mathrm{CAM}}([\Bm',\BX']), \\
\text{similarity module:}& \hspace{-1.7cm} & \hat y&=f^{\mathrm{SM}}(\Bm'',\BX'',\By),    
\end{align}
where $\Bm \in \mathbb{R}^d$ is a molecule embedding from a trainable 
or fixed molecule encoder, and $\Bm'$ and $\Bm''$ are enriched 
versions of it. Similarly, 
$\BX \in \mathbb{R}^{d \times N}$ contains the stacked 
embeddings of the support set molecules 
and $\BX'$ and $\BX''$ are their enriched versions. $\BC  \in \mathbb{R}^{d \times M} $ is a large set of stacked molecule embeddings,
$\By$ are the support set labels, and $\hat y$ is the prediction 
for the query molecule. 
Square brackets indicate concatenation, for example $[\Bm',\BX']$ is 
a matrix with $N+1$ columns. The modules $f^{\mathrm{CM}}$, 
$f^{\mathrm{CAM}}$, and $f^{\mathrm{SM}}$ are detailed in the 
paragraphs below. An overview of our architecture
is given in Figure~\ref{fig:overview_model}. The architecture also includes skip connections bypassing $f^\mathrm{CM}(.,.)$ and $f^\mathrm{CAM}(.)$ and layer normalization \citep{ba2016layer}, which are not shown in Figure\ref{fig:overview_model}.

A shared molecule encoder $f^\mathrm{ME}$ creates embeddings  
for the query molecule $\Bm=f^\mathrm{ME}(m)$, 
the support set molecules $\Bx_n=f^\mathrm{ME}(x_n)$,
and the context molecules $\Bc_m=f^\mathrm{ME}(c_m)$. 
There are many possible choices for fixed or adaptive molecule encoders 
(see Section~\ref{sec:problem}), of which we use descriptor-based
fully-connected networks because of their computational efficiency
and good accuracy \citep{dahl2014multi,mayr2016deeptox,mayr2018large}.
For notational clarity 
we denote the course of the representations 
through the architecture:
\begin{align}
    \underset{\begin{subarray}{c}\text{\sffamily symbolic~or}\\\text{\sffamily low-level~repr.}\end{subarray}} m &\overset{f^\mathrm{ME}}\longrightarrow \underset{\begin{subarray}{c}
    \text{\sffamily molecule}\\\text{\sffamily embedding}\end{subarray}}{\Bm}
    \overset{f^\mathrm{CM}}\longrightarrow \underset{\begin{subarray}{c}\text{\sffamily context}\\\text{\sffamily repr.}\end{subarray}}{\Bm'}
    \overset{f^\mathrm{CAM}}\longrightarrow \underset{\begin{subarray}{c}\text{\sffamily similarity}\\\text{\sffamily repr.}\end{subarray}}{\Bm''},\\
    \underset{\begin{subarray}{c}\text{\sffamily symbolic~or}\\\text{\sffamily low-level~repr.}\end{subarray}} {x_n} &\overset{f^\mathrm{ME}}\longrightarrow \underset{\begin{subarray}{c}\text{\sffamily molecule}\\\text{\sffamily embedding}\end{subarray}}{\Bx_n}
    \overset{f^\mathrm{CM}}\longrightarrow \underset{\begin{subarray}{c}\text{\sffamily context}\\\text{\sffamily repr.}\end{subarray}}{\Bx_n'}
    \overset{f^\mathrm{CAM}}\longrightarrow \underset{\begin{subarray}{c}\text{\sffamily similarity}\\\text{\sffamily repr.}\end{subarray}}{\Bx_n''}.
\end{align}

\subsection{Context module (CM)}
The context module associates the query and support set molecules
with a large set of context molecules, and represents them 
as weighted average of context molecule embeddings. 
The context module is realised by a continuous
Modern Hopfield Network (MHN) \citep{ramsauer2021hopfield}.
An MHN is a content-addressable
associative memory which can be built into deep 
learning architectures. There exists an analogy between the
energy update of MHNs and the attention 
mechanism of Transformers \citep{vaswani2017attention, ramsauer2021hopfield}.
MHNs are capable of storing and retrieving patterns
from a memory $\BM \in \mathbb{R}^{e \times M}$
given a state pattern $\Bxi \in \mathbb{R}^e$ 
that represents the query. 
The retrieved pattern $\Bxi^{\mathrm{new}} \in \mathbb{R}^e$ is
obtained by
\begin{align}
    \boldsymbol \xi^{\mathrm{new}} = \BM\ \Bp = \BM \ \mathrm{softmax}\left(\beta \BM^T \boldsymbol \xi \right), 
\end{align}
where $\Bp$ is called the vector of associations and 
$\beta$ is a scaling factor or inverse temperature. 
Modern Hopfield Networks have been successfully applied to 
chemistry and computational immunology \citep{seidl2022improving, widrich2020modern}.

We use this mechanism in the form of a \emph{Hopfield layer}, which 
first maps raw patterns 
to an associative space using linear transformations, 
and uses multiple simultaneous queries $\boldsymbol \Xi \in \mathbb{R}^{d \times N}$:
\begin{align}
    \label{eq:hopfield_layer}
    \mathrm{Hopfield}(\boldsymbol \BXi, \BC ) := (\BW_E \BC) \ \mathrm{softmax}\left(\beta\ (\BW_C \BC )^T \ (\BW_{\Xi} \boldsymbol \Xi) \right), 
\end{align}
where $\BW_E \in \mathbb{R}^{d \times d}$ and $\BW_{C}, \BW_{\Xi} \in \mathbb{R}^{e \times d}$ are trainable parameters
of the Hopfield layer, $\mathrm{softmax}$ is applied column-wise, and $\beta$ is a hyperparameter. 
Note that in principle the $\boldsymbol \Xi$ and 
$\BC$ could have a different second dimension as long as the linear transformations
map to the same dimension $e$.
Note that all embeddings that 
enter this module are 
first layer normalized \citep{ba2016layer}.
Several of these Hopfield layers can run in
parallel and we refer to them as "heads" in analogy to Transformers \citep{vaswani2017attention}.

The context module of our new architecture uses a Hopfield layer, where 
the query patterns are the embeddings of the 
query molecule $\Bm$ and the support set molecules $\BX$. 
The memory is composed of embeddings of 
a large set of $M$ molecules from a chemical space, 
for example reference molecules, here called context molecules $\BC$. 
Then the original embeddings $\Bm$ and $\BX$ are replaced by the 
retrieved embeddings, which are weighted averages of context molecule 
embeddings: 
\begin{align}
    \Bm'=\mathrm{Hopfield}(\Bm, \BC) \quad \text{and} \quad 
    \BX'=\mathrm{Hopfield}(\BX, \BC).
\end{align}

This retrieval step reinforces the covariance structure of the retrieved
representations (see Appendix~\ref{appsec:covariance}), which
usually enhances robustness of the models \citep{furst2021cloob} 
by removing noise.
Note that the embeddings of the query and the support set 
molecules have not yet influenced each other. 
These updated representations $\Bm', \BX'$ 
are passed to the cross-attention module.
Exemplary retrievals from the context module are included in Appendix \ref{appsec:cm_details}.

\subsection{Cross-attention module (CAM)}
For embedding-based few-shot learning methods in the field of drug discovery, 
\citet{altae2017low} showed that the representations of the molecules can be 
enriched, if the architecture allows information exchange
between query and support 
set molecules. 
\citet{altae2017low} uses an attention-enhanced LSTM variant, which updates 
the query and the support set molecule representations in an iterative fashion being aware of each other. We 
further develop this idea and combine it with the idea of using a 
transformer encoder layer \citep{vaswani2017attention} 
as a cross-attention module \citep{hou2019cross, chen2021sparse}.

The cross-attention module updates the query molecule representation $\Bm'$ 
and the support set molecule representations $\BX'$ by mutually
exchanging information, using the usual Transformer mechanism: 
\begin{align}
    [\Bm'',\BX'']=\mathrm{Hopfield}([\Bm',\BX'],  [\Bm',\BX']),
\end{align}
where $[\Bm',\BX'] \in \mathbb{R}^{d \times (N+1)}$ is the concatenation 
of the representations of the query molecule $\Bm'$ with 
the support set molecules $\BX'$ and 
we exploited that the Transformer is a special case of the Hopfield layer. 
Again, normalization is applied \citep{ba2016layer} and multiple Hopfield layers -- i.e., heads -- can run in parallel, be stacked, and equipped with skip-connections. The representations
$\Bm''$ and $\BX''$ are passed to the similarity module.

\subsection{Similarity module (SM)}
\label{sec:similarity_module}
In this module, 
pairwise similarity values $k(\Bm'', \Bx_n'')$ 
are computed between the representation of a 
query molecule $\Bm''$ 
and each molecule $\Bx_n''$ in the support set as done recently \citep{koch2015siamese,altae2017low}. 
Based on these similarity values,
the activity for the query molecule is predicted, 
building a weighted mean over the support set labels:
\begin{align}
    \hat y = \sigma\left( \tau^{-1}
    \frac{1}{N} \sum_{n=1}^{N} y_n'\ k(\Bm'', \Bx_n'') \right) ,
\end{align}
where our architecture employs dot product similarity 
of normalized representations $k(\Bm'', \Bx_n'')={\Bm''}^T \Bx''_n$.
$\sigma(.)$ is the sigmoid function 
and $\tau$ is a hyperparameter. 
Note that we use a balancing strategy 
for the labels $y_n'=\begin{cases}
N/(\sqrt{2N_A}) & \text{if}\ \ y_n=1 \\
-N/(\sqrt{2N_I}) & \, \text{else}
\end{cases}$, where $N_A$ is the number of actives and
$N_I$ is the number of inactives of the support set.

\subsection{Architecture, hyperparameter selection, and training details}
\label{sec:training_details}
\textbf{Hyperparameters.} 
The main hyperparameters of our architecture are the 
number of heads, the embedding dimension, the 
dimension of the association space of the CAM and CM, 
the learning rate schedule, the scaling parameter 
$\beta$, and the molecule encoder. 
The following hyperparameters were selected by manual
hyperparameter selection on the validation tasks. 
The molecule encoder consists of a single layer with output 
size $d=1024$ and SELU activation \citep{klambauer2017self}. 
The CM consists of one Hopfield 
layer with 8 heads. 
The dimension $e$ of the association space 
is set to 512 and $\beta=1/\sqrt{e}$. 
Since we use skip connections between all modules
the output dimension of the CM and CAM matches the input dimension. 
The CAM comprises one layer with 8 heads and an 
association-space dimension of 1088. 
For the input to the CAM, an activity encoding 
was added to the support set 
molecule representations to provide label information. 
The SM uses $\tau=32$.
For the context set, we randomly sample 
5\% from a large set of molecules -- i.e., the molecules in the FS-Mol training split --
for each batch.
For inference, we used a fixed set of 5\% of training 
set molecules as the context set for each seed.
We hypothesize that these choices about the context 
could be further improved (Section~\ref{sec:discussion}). 
We provide considered and selected hyperparameters in Appendix~\ref{appsec:mhnfs}. 

\textbf{Loss function, regularization and optimization.} 
We use the Adam optimizer \citep{kingma2014adam}
to minimize the cross-entropy loss between the predicted and known 
activity labels.  
We use a learning rate scheduler which includes a warm up phase, 
followed by a section with a constant learning rate, which is 
$0.0001$, and a third phase in which the learning rate steadily 
decreases.  As a regularization strategy, 
for the CM and the CAM a dropout rate of $0.5$ is used.
The molecule encoder has a dropout with 
rate $0.1$ for the input and $0.5$ 
elsewhere (see also Appendix~\ref{appsec:mhnfs}).

\textbf{Compute time and resources.} Training a single
\textbf{MHNfs} model on the benchmarking dataset FS-Mol takes
roughly 90 hours of wall-clock time on an A100 GPU. 
In total, roughly 15,000 GPU hours were consumed 
for this work.

\section{Related work} \label{sec:relatedwork}
Several approaches to few-shot learning in drug discovery have been suggested
\citep{altae2017low,nguyen2020meta,guo2021few,wang2021property}.
\citet{nguyen2020meta} evaluated the applicability of MAML and its variants 
to graph neural networks (GNNs) 
and \citep{guo2021few} also combine GNNs and meta-learning. 
\citet{altae2017low} suggested an approach called  
Iterative Refinement Long Short-Term Memory, in which query and support 
set embeddings can share information and update their embeddings.
Property-aware relation networks (PAR) \citep{wang2021property} use
an attention mechanism to enrich representations from cluster centers
and then learn a relation graph between molecules. 
\citet{chen2022meta} propose
to adaptively learn kernels and apply their method to few-shot 
drug discovery with predictive performance for larger 
support set sizes. 
Recently, \citet{stanley2021fs} generated a benchmark dataset 
for few-shot learning methods in drug discovery and 
provided some baseline results.

Many successful
deep neural network architectures 
use external memories, 
such as the neural Turing machine \citep{graves2014neural},
memory networks \citep{weston2014memory},
end-to-end memory networks \citep{sukhbaatar2015end}.
Recently, the connection between continuous modern 
Hopfield networks \citep{ramsauer2021hopfield}, 
which are content-addressable associative memories,
and Transformer architectures \citep{vaswani2017attention} 
has been established. We refer to \citet{le2021memory} for 
an extensive overview of memory-based architectures. 
Architectures with external memories have also been used 
for meta-learning \citep{vinyals2016matching,santoro2016meta} and 
few-shot learning \citep{munkhdalai2017meta, ramalho2018adaptive,ma2021few}.

\section{Experiments}
\label{sec:experiments}
\subsection{Benchmarking on  FS-Mol}
\label{fsmol_exp_setup}

\textbf{Experimental setup.} Recently, the dataset FS-Mol \citep{stanley2021fs}
was proposed to benchmark few-shot learning methods in drug discovery.
It was extracted from ChEMBL27 and comprises in total 489,133 measurements, 
233,786 compounds and 5,120 tasks. 
Per task, the mean number of data points is 94. 
The dataset is well balanced as the mean ratio of active and inactive molecules is close to 1.
The FS-Mol benchmark dataset defines 4,938 training, 40 validation and 157 test tasks, 
guaranteeing disjoint task sets. 
\citet{stanley2021fs} precomputed extended connectivity fingerprints (ECFP) \citep{rogers2010extended} 
and key molecular physical descriptors, which were defined by RDKit \citep{landrum2006rdkit}. 
While methods would be allowed to use other 
representations of the input molecules, such as 
the molecular graph, 
\textcolor{black}{we used a concatenation of these ECFPs and RDKit-based descriptors.}
We use the main benchmark setting of FS-Mol with 
support set size 16, which is close to the 5- and 10-shot 
settings in computer vision, 
and stratified random split \citep[Table 2]{stanley2021fs}
for a fair method comparison (see also Section~\ref{appsec:sss}).  

\begin{table}
    \centering
    \begin{threeparttable}[b]
      \caption{Results on FS-MOL [$\Delta \text{AUC-PR}$]. The best method is marked bold. Error bars represent standard errors across tasks according to \citet{stanley2021fs}. The metrics are also averaged 
      across five training reruns and ten draws of support sets. In brackets the number of tasks per category is reported. \label{tab:results_fsm}
      }
      \begin{tabular}{lcccc}
        \toprule
        Method & All [157] & Kin. [125] & Hydrol. [20] & Oxid.[7]  \\ \midrule
        GNN-ST\tnote{a}\hspace{0.5em}\citep{stanley2021fs}         &  .029 $\pm$ .004 & .027 $\pm$ .004 & .040 $\pm$ .018 & .020 $\pm$ .016\\
        MAT\tnote{a}\hspace{0.5em}\citep{maziarka2020molecule}            & .052 $\pm$ .005 & .043 $\pm$ .005 & .095 $\pm$ .019 & .062 $\pm$ .024\\
        Random Forest\tnote{a}\hspace{0.5em}\citep{breiman2001random}  & .092 $\pm$ .007 & .081 $\pm$ .009 & .158 $\pm$ .028 & .080 $\pm$ .029\\
        GNN-MT\tnote{a}\hspace{0.5em}\citep{stanley2021fs}         & .093 $\pm$ .006 & .093 $\pm$ .006 & .108 $\pm$ .025 & .053 $\pm$ .018\\
        Similarity Search          & .118 $\pm$ .008 & .109 $\pm$ .008 & .166 $\pm$ .029 & .097 $\pm$ .033  \\
        GNN-MAML\tnote{a}\hspace{0.5em}\citep{stanley2021fs}       & .159 $\pm$ .009 & .177 $\pm$ .009  & .105 $\pm$ .024 & .054 $\pm$ .028 \\
        {PAR} \citep{wang2021property} & {.164 $\pm$ .008} & {.182 $\pm$ .009} & {.109 $\pm$ .020} & {.039 $\pm$ .008}  \\
        Frequent Hitters & .182 $\pm$ .010 & .207 $\pm$ .009 & .098 $\pm$ .009 & .041 $\pm$ .005  \\
        ProtoNet\tnote{a}\hspace{0.5em}\citep{snell2017prototypical}       & .207 $\pm$ .008 & .215 $\pm$ .009 & .209 $\pm$ .030 & .095 $\pm$ .029\\
        Siamese Networks \citep{koch2015siamese} & .223 $\pm$ .010 & .241 $\pm$ .010 & .178 $\pm$ .026 & .082 $\pm$ .025\\
        IterRefLSTM\hspace{0.5em}\citep{altae2017low} & .234 $\pm$ .010 & .251 $\pm$ .010 & .199 $\pm$ .026 & .098 $\pm$ .027  \\
        {ADKF-IFT\tnote{b}\hspace{0.5em}\citep{chen2022meta} } & {.234 $\pm$ .009} & {.248 $\pm$ .020} & {\textbf{.217} $\pm$ .017} & {\textbf{.106} $\pm$ .008}  \\
        MHNfs (ours) & \textbf{.241} $\pm$ .009 & \textbf{.259} $\pm$ .010 & .199 $\pm$ .027 & .096 $\pm$ .019  \\
        \bottomrule
      \end{tabular}
      \begin{tablenotes}
        \item [a] metrics from \citet{stanley2021fs}.
        \item [b] { results from \citet{chen2022meta}.}
      \end{tablenotes}
    \end{threeparttable}
\end{table}

\textbf{Methods compared.}
Baselines for few-shot learning and our proposed method \textbf{MHNfs} were compared against each other.
The \textbf{Frequent Hitters} model is a naive baseline that 
ignores the provided support set and 
therefore has to learn to predict the average activity of a molecule.
This method can potentially discriminate so-called frequent-hitter molecules \citep{stork2019hit} against
molecules that are inactive across many tasks. 
We also added \textbf{Similarity Search} \citep{cereto2015molecular} as a baseline. Similarity search is 
a standard chemoinformatics technique, used in situations with  single or few known actives. 
In the simplest case, the search finds similar molecules by computing 
a fingerprint or descriptor-representation of the molecules and using a similarity measure $k(.,.)$ --
such as Tanimoto Similarity \citep{tanimoto1960ibm}. 
Thus, Similarity Search, as used in chemoinformatics, 
can be formally written as $\hat y = 1/N \sum_{n=1}^N  y_n\ k(\Bm, \Bx_n),$ 
where $\Bx_1,\ldots,\Bx_{\mathrm{n}}$ come from a fixed molecule encoder, 
such as chemical fingerprint or descriptor calculation. A natural extension of Similarity Search with fixed chemical 
descriptors is \textbf{Neural Similarity Search or Siamese networks} 
\citep{koch2015siamese}, which extend the classic 
similarity search by learning a molecule encoder: 
$\hat y = \sigma\left( \tau^{-1} \frac{1}{N} \sum_{n=1}^{N} y_n' \ f^{\mathrm{ME}}_{\Bw}(\Bm)^{T} \ f^{\mathrm{ME}}_{\Bw}(\Bx_{n}) \right)$.
Furthermore, we re-implemented the \textbf{IterRefLSTM} \citep{altae2017low} in PyTorch.
The \textbf{IterRefLSTM} model consists of three modules. First, a molecule encoder 
maps the query and support set molecules to its representations $\Bm$ and $\BX$. 
Second, an attention-enhanced LSTM variant, the actual \textbf{IterRefLSTM}, iteratively updates 
the query and support set molecules, enabling information sharing between the 
molecules: $[\Bm', \BX'] = \mathrm{IterRefLSTM}_L([\Bm, \BX])$, where the 
hyperparameter $L$ controls the number of iteration steps of the \textbf{IterRefLSTM}. Third, a similarity module computes attention weights based on the representations: $\Ba = \mathrm{softmax}\left(k \left(\Bm', \BX'\right)\right)$.
These representations are then used for the final prediction: $\hat y = \sum_{i=1}^{N} a_i y_i$. For further details, see
Appendix~\ref{appsec:iterreflstm}. 
The \textbf{Random Forest} baseline uses the chemical descriptors and is trained in standard supervised manner 
on the support set molecules for each task.
The method \textbf{GNN-ST} is a graph neural network \citep{stanley2021fs,gilmer2017neural} that
is trained from scratch for each task. 
The \textbf{GNN-MT} uses a two step strategy: First, the model is pretrained on a large dataset on related \textcolor{black}{tasks}; 
second, an output layer is constructed to the few-shot task via linear probing \citep{stanley2021fs,alain2016understanding}.
The \textbf{Molecule Attention Transformer (MAT)} is pre-trained in a 
self-supervised fashion and fine-tuning is performed for the few-shot task \citep{maziarka2020molecule}.
\textbf{GNN-MAML} is based on MAML \citep{finn2017model}, 
and uses a model-agnostic meta-learning strategy 
to find a general core model from which one can easily adapt to single tasks.
Notably, GNN-MAML also can be seen as a proxy for \textbf{Meta-MGNN} \citep{guo2021few}, which enriches the gradient update step in the outer loop of the MAML-framework by an attention mechanism and uses an additional atom type prediction loss and a bond reconstruction loss.
\textbf{ProtoNet} \citep{snell2017prototypical} includes a molecule encoder, 
which maps query and support set molecules to representations in an embedding space. 
In this embedding space, prototypical representations of each class are built 
by taking the mean across all related support set molecules for each class (details in Appendix~\ref{appsec:protonet}). 
The \textbf{PAR} model \citep{wang2021property} includes a GNN which creates initial molecule embeddings.
These molecule embeddings are then enriched by an attention mechanism. Finally, another GNN 
learns relations between support and query set molecules.
\textcolor{black}{The \textbf{PAR} model has shown good results for datasets which just include very few tasks such as Tox21 \citep{wang2021property}.}
\citet{chen2022meta} suggest a framework for learning deep kernels by interpolating between meta-learning and conventional deep kernels, which results in the \textbf{ADKF-IFT} model. \textcolor{black}{The model has exhibited especially high performance for large support set sizes}.
For all methods the most important hyperparameters
were adjusted on the validation tasks of FS-Mol.

\textbf{Training and evaluation.}
For the model implementations, we used PyTorch \citep[BSD license]{paszke2017automatic}. We used PyTorch Lightning \citep[Apache 2.0 license]{falcon2019pytorch} as a framework for training and test logic, hydra for config file handling \citep[Apache 2.0 license]{Yadan2019Hydra} and  Weights~\&~Biases \citep[MIT license]{wandb} as an experiment tracking tool. 
We performed five training reruns with different seeds for all methods, except Classic Similarity Search as there is no variability across seeds. 
Each model was evaluated ten times by drawing support sets with ten different seeds.

\textbf{Results.}
The results in terms of area under precision-recall curve (AUC-PR) are presented 
in Table~\ref{tab:results_fsm}, where the difference to a random classifier is reported
($\Delta$AUC-PR). The standard error is reported across tasks. 
Surprisingly, the naive baseline \textbf{Frequent Hitters},
that neglects the support set,
has outperformed most of the few-shot learning methods, 
except for the embedding based methods and ADKF-IFT.
\textbf{MHNfs} has outperformed all other methods with respect 
to $\Delta$AUC-PR across all tasks, 
including the \textbf{IterRefLSTM} model
($p$-value 1.72e-7, paired Wilcoxon test), \textcolor{black}{the \textbf{ADKF-IFT} model ($p$-value <1.0e-8, Wilcoxon test), and the \textbf{PAR} model ($p$-value <1.0e-8, paired Wilcoxon test)}.

\begin{figure}
    \centering
    \includegraphics[width=0.48\textwidth]{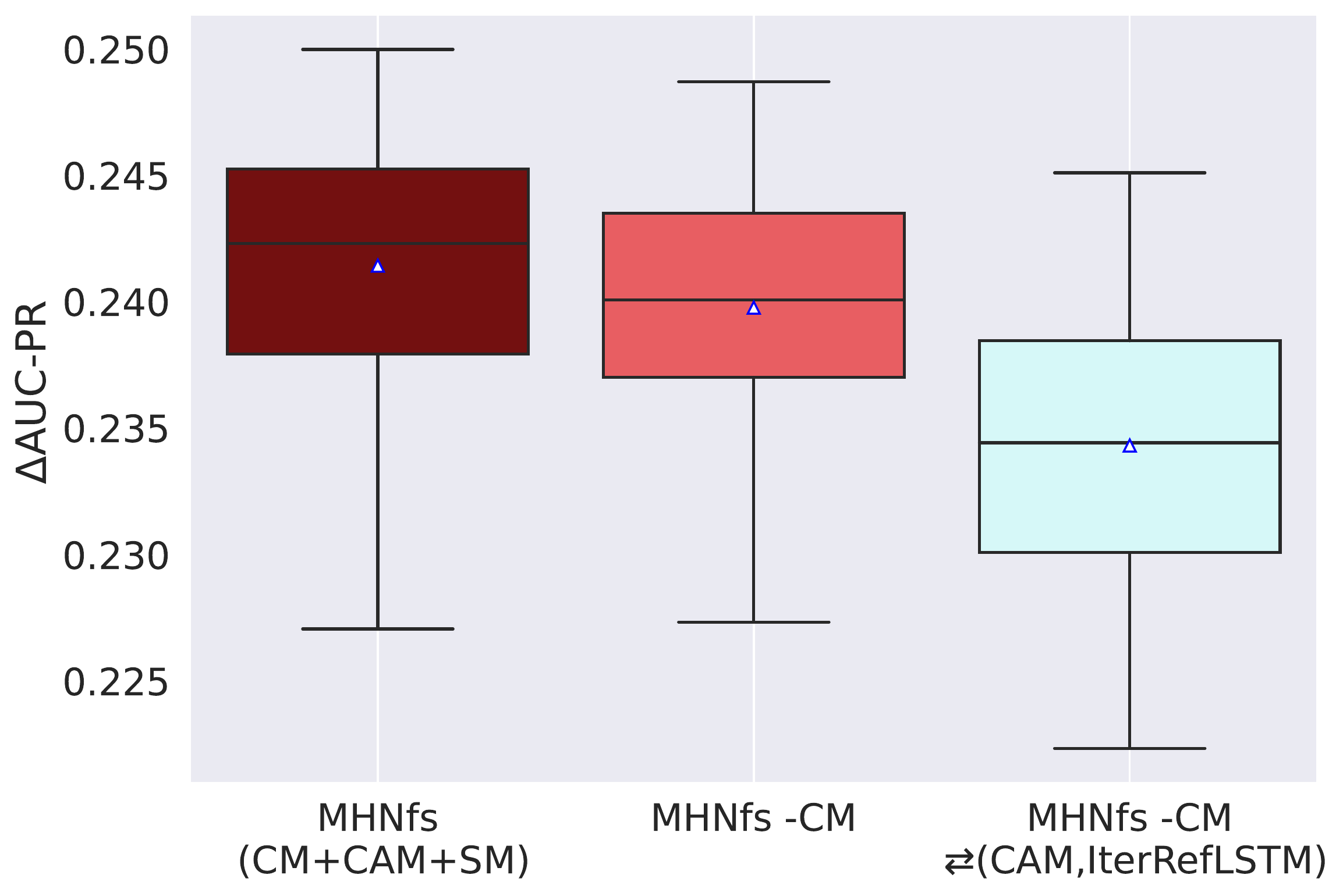}
    \hspace{0.1cm}
    \includegraphics[width=0.48\textwidth]{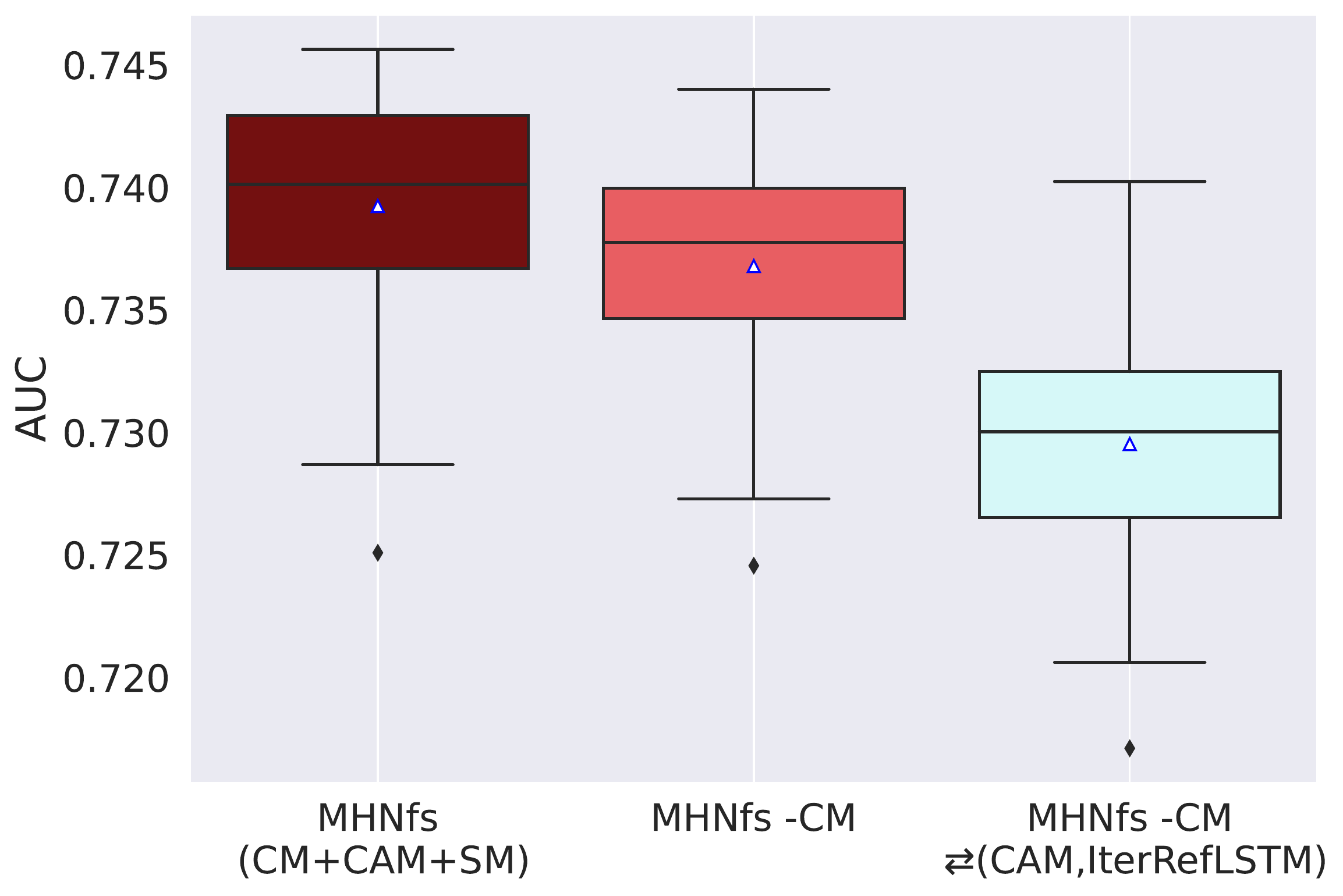}
    \caption{Results of the ablation study. The boxes show the median, mean 
    and the variability of the average predictive performance of the methods
    across training reruns and draws of support sets. The performance significantly drops when the context module is removed (light red bars), and when additionally the cross-attention module is replaced with the \textbf{IterRefLSTM} 
    module (light blue bars). This indicates that our two newly
    introduced modules, CM and CAM,
    play a crucial role in MHNfs.
    }
    \label{fig:ablation_study}
\end{figure}

\subsection{Ablation study}
\textbf{MHNfs} has two new main components
compared
to the most similar previous state-of-the-art method \textbf{IterRefLSTM}:
i) the context module, and ii) the cross-attention
module which replaces
the LSTM-like module. 
To assess the effects of these
components,
we performed an ablation study. 
Therefore, we compared \textbf{MHNfs} to a method 
that does not have the context module ("MHNfs -CM") and
to a method that does not have the context module 
and uses an LSTM-like module instead of the CAM ("MHNfs -CM $\rightleftharpoons$(CAM,IterRefLSTM)").
For the ablation study, we used all 5 training reruns and evaluated 10 times on the test set with different support sets.
The results of this ablation steps are presented
in Figure~\ref{fig:ablation_study}. 
Both removing the CM and exchanging the CAM 
with the \textbf{IterRefLSTM} module were detrimental for 
the performance of the method ($p$-value 0.002 and
$1.72$e$-7$, respectively; paired Wilcoxon test). The
difference was even more pronounced under domain 
shift (see Appendix~\ref{appsec:domain_shift_tox21}). 
Appendix~\ref{appsec:domain_shift_alldesign} contains
a second ablation study
that examines the overall effects 
of the context, 
the cross-attention, 
the similarity module,
and the molecule encoder of \textbf{MHNfs}.

\subsection{Domain shift experiment}
\label{sec:domain-shift}

\textbf{Experimental setup.}
The Tox21 dataset consists of 12,707 chemical compounds, for which measurements for up to 12 different toxic effects are reported \citep{mayr2016deeptox, huang2016tox21challenge}. It was published with a fixed training, validation and test split. 
\textcolor{black}{
State-of-the-art supervised learning methods that have access to the full training set reach AUC performance values between $0.845$ and $0.871$ \citep{klambauer2017self, duvenaud2015convolutional, li2017learning, li2021trimnet, zaslavskiy2019toxicblend, alperstein2019all}.
}
For our evaluation, we re-cast Tox21 
as a few-shot learning setting and draw small support sets from 
the 12 tasks. The compared methods were pre-trained on FS-Mol and obtain small support sets from Tox21. Based on the 
support sets, the methods had to predict the activities of 
the Tox21 test set. Note that there is a strong domain shift 
from drug-like molecules of FS-Mol to environmental chemicals, pesticides,
food additives of Tox21. The domain shift also concerns the outputs
where a shift from kinases, hydrolases, and oxidoreductases of FS-Mol
to nuclear receptors and stress responses of Tox21 is present. 

\textbf{Methods compared.} We compared \textbf{MHNfs}, 
the runner-up method \textbf{IterRefLSTM},
and \textbf{Similarity Search} -- since it has been 
widely used for such purposes for decades \citep{cereto2015molecular}. 

\textbf{Training and evaluation.}
We followed the procedure of \citet{stanley2021fs}
for data-cleaning, preprocessing and extraction
of the fingerprints and descriptors used in FS-Mol. 
After running the cleanup step, 
8,423 molecules remained for the domain shift experiments. 
From the training set, 8 active and 8 inactive molecules 
per task were randomly selected to build the 
support set. 
The test set molecules were used as query molecules. 
The validation set molecules were not used at all.
During test-time, a support set 
was drawn ten times for each task. Then, the 
performance of the models were evaluated for these support sets, 
using the 
area under precision-recall curve (AUC-PR), 
analogously to the FS-Mol benchmarking 
experiment reported as the difference to a random classifier 
($\Delta$AUC-PR), 
and the area under receiver operating characteristic curve (AUC) 
metrics. 
The performance values report the mean over all combinations regarding the 
training reruns and the support set sampling iterations. 
Error bars indicate the standard deviation.

\textbf{Results.}
The Hopfield-based context retrieval method has significantly 
outperformed the IterRefLSTM-based model 
($p_{\Delta\mathrm{AUC-PR}}$-value $3.4$e$-5$, 
$p_{\mathrm{AUC}}$-value $2.5$e-$6$, paired Wilcoxon 
test) and the Classic Similarity Search 
($p_{\Delta\mathrm{AUC-PR}}$-value $2.4$e-$9$, 
$p_{\mathrm{AUC}}$-value $7.6$e-$10$, paired Wilcoxon 
test) and therefore showed robust performance on the toxicity 
domain, see Table~\ref{tab:results_tox21}. 
Notably, all models were trained on the FS-Mol dataset and then applied to the Tox21 dataset without adjusting any weight parameter.

\begin{table}
    \centering
    \begin{threeparttable}[b]
      \caption{Results of the domain shift experiment on Tox21 [AUC, $\Delta \text{AUC-PR}$]. The best method is marked bold. Error bars represent standard deviation across training reruns and draws of support sets \label{tab:results_tox21}}
      \begin{tabular}{lcccc}
        \toprule
        Method & AUC & $\Delta \text{AUC-PR}$  \\ \midrule
        Similarity Search (baseline) &  .629 $\pm$ .015 & .061 $\pm$ .008\\
        IterRefLSTM\hspace{0.5em}\citep{altae2017low} &  .664 $\pm$ .018 & .067 $\pm$ .008\\
        MHNfs (ours)  &  \textbf{.679} $\pm$ .018 & \textbf{.073} $\pm$ .008\\
        \bottomrule
      \end{tabular}
    \end{threeparttable}
\end{table}

\section{Conclusion}
\label{sec:discussion}
\label{sec:broader_impact}
We have introduced a new architecture for few-shot learning in drug
discovery, namely MHNfs, that is based on the novel concept to enrich 
molecule representations with context.
In a benchmarking experiment the architecture outperformed all other methods and in a domain shift study the robustness and transferability has been assessed.
We envision that the context module can be applied to many different areas, enriching learned representations analogously to our work. For discussion, see \ref{appsec:discussion}.

\section*{Acknowledgements}
The ELLIS Unit Linz, the LIT AI Lab, the Institute for Machine Learning, are supported by the Federal State Upper Austria. IARAI is supported by Here Technologies. 
We thank Merck Healthcare KGaA for the collaboration.
Further, we thank the projects AI-MOTION (LIT-2018-6-YOU-212), DeepFlood (LIT-2019-8-YOU-213), Medical Cognitive Computing Center (MC3), INCONTROL-RL (FFG-881064), PRIMAL (FFG-873979), S3AI (FFG-872172), DL for GranularFlow (FFG-871302), EPILEPSIA (FFG-892171), AIRI FG 9-N (FWF-36284, FWF-36235), ELISE (H2020-ICT-2019-3 ID: 951847), Stars4Waters (HORIZON-CL6-2021-CLIMATE-01-01). We thank Audi.JKU Deep Learning Center, TGW LOGISTICS GROUP GMBH, Silicon Austria Labs (SAL), FILL Gesellschaft mbH, Anyline GmbH, Google, ZF Friedrichshafen AG, Robert Bosch GmbH, UCB Biopharma SRL, Verbund AG, GLS (Univ. Waterloo) Software Competence Center Hagenberg GmbH, T\"{U}V Austria, Frauscher Sensonic and the NVIDIA Corporation.

\bibliography{bib_my,bib}
\bibliographystyle{apalike}

\newpage

\appendix
\newpage

\setcounter{theorem}{0}
\setcounter{definition}{0}
\setcounter{table}{0}
\setcounter{figure}{0}
\setcounter{equation}{0}

\renewcommand{\thefigure}{A\arabic{figure}}
\renewcommand{\thetable}{A\arabic{table}}
\renewcommand{\theequation}{A\arabic{equation}}

\resumetocwriting

\section{Appendix}

\renewcommand{\contentsname}{\normalsize Contents of the appendix}
{
  \hypersetup{linkcolor=black}
    \tableofcontents
}

\subsection{Details on methods} 
    Few-shot learning methods in drug discovery 
    can be described 
    as models with adaptive parameters $\Bw$ that 
    use a support set $\BZ=\{(\Bx_1,y_1),\ldots,(\Bx_N,y_N) \}$ 
    \footnote{We use $\BZ$ to denote the support set 
    of already embedded molecules to keep the 
    notation uncluttered. More correctly, the 
    methods have access to the raw support set $Z=\{(x_1,y_1),\ldots,(x_N,y_N)\}$,
    where $x_n$ is a symbolic, such as the molecular graph,
    or low-level representation of the molecule.}
    as additional input
    to predict a label $\hat y$ for a molecule $\Bm$
    \begin{align}
        \hat y = g_{\Bw}(\Bm,\BZ).
    \end{align}
    
    Optimization-based methods, such as MAML \citep{finn2017model}, use the support set 
    to update the parameters $\Bw$
    \begin{align}
        \hat y = g_{ a(\Bw;\BZ) }(\Bm),
    \end{align}
    where $a(.)$ is a function that adapts $\Bw$ of $g$
    based on $\BZ$ for example via gradient-descent. 
    
    Embedding-based methods use a different approach 
    and learn representations of the support set 
    molecules $\{\Bx_1,\ldots,\Bx_N\}$, sometimes written 
    as stacked embeddings $\BX \in \mathbb{R}^{d \times N}$, 
    and the query molecule $\Bm$, and
    some function that associates these two types
    of information with each other. We describe
    the embedding-based methods
    Similarity Search in Section~\ref{appsec:simsearch},
    Neural Similarity Search in Section~\ref{appsec:neuralsim},
    ProtoNet in Section~\ref{appsec:protonet},
    IterRefLSTM in Section~\ref{appsec:iterreflstm},
    PAR in Section~\ref{appsec:PAR},
    and MHNfs in the main paper and details 
    in Section~\ref{appsec:mhnfs}. The "frequent hitters" 
    baseline is described in Section~\ref{appsec:frequent_hitters}.

\subsubsection{Frequent hitters: details and hyperparameters}
    \label{appsec:frequent_hitters}
    The "frequent hitters" model $g^{\mathrm{FH}}$ is a baseline 
    that we implemented and included in the method comparison. 
    This method uses the usual training scheme of sampling a query molecule $\Bm$ with a label $y$, having access to a support set $\BZ$.
    In contrast to the usual models of the type $g_{\Bw}(\Bm, \BZ)$, 
    the frequent hitters model $g^{\mathrm{FH}}$ neglects the support set:
    
     \begin{align}
        \label{eq:frequent_hitters}
            \hat y = g^{\mathrm{FH}}_{\Bw}(\Bm).
        \end{align}
        
    Thus, during training for the same molecule $\Bm$, the model 
    might have to predict both $y=1$ and $y=-1$, since the molecule 
    can be active in one task and inactive in another task. Therefore,
    the model tends to predict average activity of a molecule to 
    minimize the cross-entropy loss. We chose an additive combination of the \textcolor{black}{Morgan fingerprints, RDKit fingerprints, and MACCS keys} for the input representation to the MLP. 
    
    \paragraph{Hyperparameter search.}
    We performed manual hyperparameter search on the validation set and report the explored hyperparameter space (Table \ref{tab:hyperparameterSpace_FH}). We use early-stopping based on validation average-precision, a patience of 3 epochs and train for a maximum of 20 epochs with a linear warm-up learning-rate schedule for the first 3 epochs. 
    
    \begin{table}
        \centering
        \begin{threeparttable}[b]
        \caption{Hyperparameter space considered for the Frequent Hitters model. The hyperparameters of the best configuration are marked bold. }
        \label{tab:hyperparameterSpace_FH}
        \begin{tabular}{ll}
        \toprule Hyperparameter & Explored values \\
        \midrule
            Number of hidden layers & 1, \textbf{2}, 4  \\
            Number of units per hidden layer & 1024, \textbf{2048}, 4096 \\
            Output dimension & \textbf{512}, 1024 \\
            Activation function & \textbf{ReLU} \\
            Learning rate & \textbf{0.0001}, 0.001 \\
            Optimizer & Adam,\textbf{AdamW} \\
            Weight decay & 0, {$ \boldsymbol{ 0.01} $} \\
            Batch size & 32, 128, 512, 2048, \textbf{4096} \\
            Input Dropout & 0,  \textbf{0.1} \\
            Dropout & 0.1, 0.2, 0.3, \textbf{0.4}, 0.5 \\
            Layer-normalization & False, \textbf{True} \\
            ~~\textbullet~ Affine & \textbf{False}, True \\
            Similarity function & \textbf{dot product} \\ 
        \bottomrule    
        \end{tabular}
        \end{threeparttable}
    \end{table}

\subsubsection{Classic similarity search: details and hyperparameters}
\label{appsec:simsearch}
    Similarity Search \citep{cereto2015molecular} 
    is a classic chemoinformatics technique used in situations
    in which a single or few actives are known. In the simplest case, 
    molecules that are similar to a given active molecule are searched by computing a fingerprint or descriptor-representation $f^{\mathrm{desc}}(m)$ of the molecules and using a similarity measure $k(.,.)$, such as Tanimoto Similarity\citep{tanimoto1960ibm}. Thus, the Similarity Search as used in chemoinformatics can be formally written as: 
    \begin{align}
        \hat y = 1/N \sum_{n=1}^N  y_n\ k( f^{\mathrm{desc}} (m), f^{\mathrm{desc}}(x_n) ), 
    \end{align}
    where the function $f^{\mathrm{desc}}$ maps the molecule to its chemical 
    descriptors or fingerprints
    and takes the role of both
    the molecule encoder
    and the support set encoder. 
    Then, an association function, consisting of 
    a) the similarity measure $k(.,.)$ and b) mean pooling across
    molecules weighted by their similarity and activity, is used to compute the predictions. 
    
    Notably, there are many variants of Similarity Search \citep{cereto2015molecular,wang2010application,eckert2007molecular,geppert2008support,willett2014calculation,sheridan2002we,riniker2013open} of which some correspond to recent 
    few-shot learning methods with a fixed molecule encoder. For example, 
    \citep{geppert2008support} suggest to use centroid molecules, i.e., prototypes
    or averages of active molecules. This is equivalent to the idea of Prototypical Networks \citep{snell2017prototypical}. \citet{riniker2013open} are aware of 
    different fusion strategies for sets of active or inactive molecules, which 
    corresponds to different pooling strategies of the support set. 
    Overall, the variants of the classic Similarity Search are highly similar to embedding-based few-shot
    learning methods except that they have a fixed instead of a learned molecule encoder. 
    
    \paragraph{Hyperparameter search.}
    For the Similarity Search, there were two decisions to make which was firstly the similarity metric and secondly the question whether we should use a balancing strategy like shown in Section \ref{sec:similarity_module}. We decided for the dot-product as a similarity metric and for using the balancing strategy.
    These decisions were made by evaluating the models on the validation set.

\subsubsection{Neural Similarity Search or Siamese networks: details and hyperparameters}
\label{sssec:detailsNeuralSimSearch}
\label{appsec:neuralsim}
    If the fixed encoder $f^{\mathrm{desc}}$ of the Classic Similarity Search is replaced by learned encoders $f^{\mathrm{ME}}_{\Bw}$, Neural Similarity Search variants naturally arise.

\begin{figure}
        \centering
        \includegraphics[width=1\textwidth]{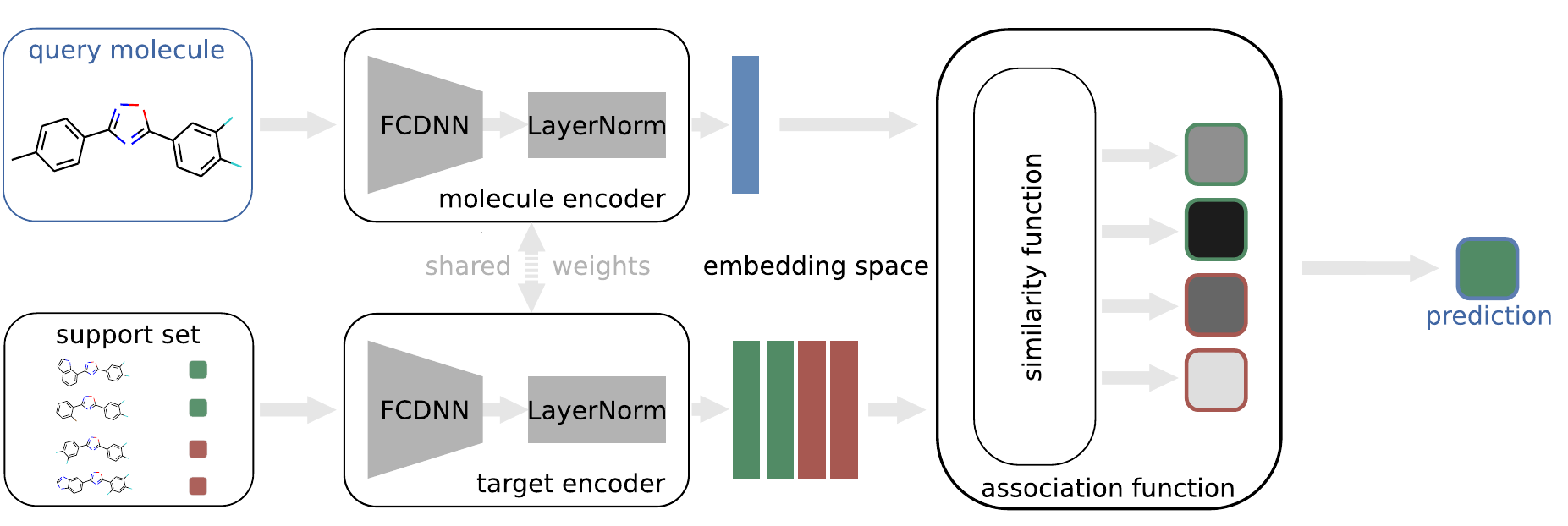}
        \caption{Schematic overview of the implemented Neural Similarity Search variant}
        \label{fig:overview_neuralSearch}
    \end{figure}

    A lot of related work already was done \citep{koch2015siamese, hertz2006learning, ye2018deep, torres2020exploring}. We adapted these ideas, such that a fully-connected deep neural network followed by a Layer Normalization \citep{ba2016layer} operation, $f^{\mathrm{ME}}_{\Bw}$
    with adaptive parameters $\Bw$, is used in a Siamese fashion to compute the embeddings for the input molecule and the support set molecules.
    Within an association function block, pairwise similarity values for the query molecule and each support set molecule are computed, associating both embeddings via the dot product. Based on these similarity values, the activity for the query molecule is predicted, building the weighted mean over the support set molecule labels:
    \begin{align}
        \hat y = \sigma\left( \tau^{-1} 
        \frac{1}{N} \sum_{n=1}^{N} y_n' \ f^{\mathrm{ME}}(m)^{T} \ f^{\mathrm{ME}}(x_{n}) \right) ,
    \end{align}
    where  $\sigma(.)$ is the sigmoid function 
    and $\tau$ is a hyperparameter in the range 
    of $\sqrt{d}$. 
    Note that this method uses a balancing strategy 
    for the labels $y_n'=\begin{cases}
    N/(\sqrt{2N_A}) & \text{if}\ \ y_n=1 \\
    -N/(\sqrt{2N_I}) & \, \text{else}
    \end{cases}$, where $N_A$ is the number of actives and
    $N_I$ is the number of inactives of the support set.
Figure~ \ref{fig:overview_neuralSearch} provides an schematic overview of the Neural Similarity Search variant.

    We trained the 
    model using the Adam optimizer \citep{kingma2014adam} to minimize binary cross-entropy loss. 

    \paragraph{Hyperparameter search.}
    We performed manual hyperparameter search on the validation set. 
    We report the explored hyperparameter space (Table \ref{tab:hyperparameterSpace}). 
    Bold values indicate the selected hyperparameters for the final model.
    
    \begin{table}
        \centering
        
        \begin{threeparttable}
         \caption{Hyperparameter space considered for the Neural Similarity Search model selection. The hyperparameters of the best configuration
        are marked bold. }
        \label{tab:hyperparameterSpace}
        \begin{tabular}{ll}
        \toprule Hyperparameter & Explored values \\
        \midrule
            Number of hidden layers & 1, \textbf{2}, 4  \\
            Number of units per hidden layer & \textbf{1024}, 4096 \\
            Output dimension & \textbf{512}, 1024 \\
            Activation function & ReLU, \textbf{SELU}\\
            Learning rate & 0.0001, \textbf{0.001}, 0.01\\
            Optimizer & \textbf{Adam}\\
            Weight decay & \textbf{0}, $1\cdot10^{-4}$ \\
            Batch size & \textbf{4096} \\
            Input Dropout & \textbf{0.1} \\
            Dropout & \textbf{0.5} \\
            Layer-normalization & False, \textbf{True} \\
            ~~\textbullet~ Affine & \textbf{False} \\
            Similarity function & cosine similarity, \textbf{dot product}, MinMax similarity\\ 
        \bottomrule    
        \end{tabular}
        \end{threeparttable}
    \end{table}
   
\subsubsection{ProtoNet: details and hyperparameters}
\label{appsec:protonet}
    Prototypical Networks (ProtoNet) \citep{snell2017prototypical} 
    learn a prototype $\Br$ for each class. 
    Concretely, the support set $Z$
    is class-wise separated into $Z^{+} \coloneqq \{ (x, y) \in Z \mid y = 1 \}$ 
    and $Z^{-} \coloneqq \{ (x, y) \in Z \mid y = -1 \}$. For the subsets $Z^{+}$ 
    and $Z^{-}$ prototypical representations $\Br^{+}$ and $\Br^{-}$ 
    can be computed by 
    \begin{align}
        \Br^{+} = \frac{1}{|Z^{+}|} \cdot \sum_{(x, y) \in Z^{+}} f^{\mathrm{ME}}(x)
    \end{align}
    and 
    \begin{align}
        \Br^{-} = \frac{1}{|Z^{-}|} \cdot \sum_{(x, y) \in Z^{-}} f^{\mathrm{ME}}(x).
    \end{align}
    The prototypical representations $\Br^{+}, \Br^{-} \in \mathbb{R}^d$ and the query molecule embedding $\Bm \in \mathbb{R}^d$ 
    are then used 
    to make the final prediction: 
    \begin{align}
    \hat y = \frac{ \exp({-\Bd(\Bm, \Br^{+})}) }{ \exp({-\Bd(\Bm, \Br^{+})}) + \exp({-\Bd(\Bm, \Br^{-})})}, 
    \end{align}
    where $\Bd$ is a distance metric.
    
   \paragraph{Hyperparameter search.}
    Hyperparameter search has been done by \citet{stanley2021fs}, 
    to which we refer here. 
    ECFP fingerprints and descriptors created by a GNN, which operates on the molecular graph, are fed into a fully connected neural network, which maps the input into an embedding space with the dimension of 512.
   \citet{stanley2021fs} use the Mahalanobis distance to measure the similarity between a query molecule and the prototypical representations in the embedding space.
   The learning rate is $0.001$ and the batch size is 256. 
   The implementation can be found here \url{https://github.com/microsoft/FS-Mol/blob/main/fs_mol/protonet_train.py} and important hyperparameters are chosen here \url{https://github.com/microsoft/FS-Mol/blob/main/fs_mol/utils/protonet_utils.py}.
   
   \paragraph{Connection to Siamese networks and 
   contrastive learning with InfoNCE.} 
   If instead of the negative distance $-\Bd(.,.)$
   the dot product similarity measure  
   with appropriate scaling is used, ProtoNet
   for two classes becomes equivalent to Siamese Networks.
   Note that in our study, another difference is that 
   ProtoNet uses a GNN for the encoder, whereas the encoder of the Siamese Networks
   is a descriptor-based fully-connected network. In case of dot product as similarity measure, 
   the objective also becomes
   equivalent to contrastive learning with 
   the InfoNCE objective \citep{oord2018representation}.

\subsubsection{IterRefLSTM: details and hyperparameters}
\label{appsec:iterreflstm}
    \citet{altae2017low} modified the idea of Matching Networks \citep{vinyals2016matching}
    by replacing the LSTM with their Iterative Refinement 
    Long Short-Term Memory (IterRefLSTM). The use of the IterRefLSTM
    empowers the architecture to update not only the 
    embeddings for the query molecule but also adjust the 
    representations of the support set molecules.
    
    For the IterRefLSTM model, query molecule embedding $\Bm= f^{\mathrm{ME}}_{\Bth_1}(m)$ 
    and support set molecule embeddings $\Bx_n= f^{\mathrm{ME}}_{\Bth_2}(x_n)$ are created using 
    two potentially different molecule encoders for the
    query molecule $m$ and the support set molecules 
    $x_1,\ldots,x_N$.
    The query and support set molecule embeddings are then updated by an LSTM-like module -- the actual IterRefLSTM:
    \begin{align*}
        [\Bm',\BX'] &= \mathrm{IterRefLSTM}_{L}( [\Bm, \BX]).
    \end{align*}
    Here, $\Bm'$ and $\BX'$ contain the updated representations for the 
    query molecule and the support set molecules. The IterRefLSTM denotes the function 
    which updates these representations. 
    The main property of the IterRefLSTM module is that it 
    is permutation-equivariant, thus a permutation $\pi(.)$
    of the input elements results in the permutation 
    of output elements: $\pi([\Bm',\BX']) = \mathrm{IterRefLSTM}_{L}( \pi([\Bm, \BX]))$. Therefore, the full architecture is invariant
    to permutations of the support set elements. 
    For details, we refer to 
    \citet{altae2017low}. 
    The hyperparameter $L \in \mathbb N$ controls the number of 
    iteration 
    steps of the IterRefLSTM.

    The IterRefLSTM also includes a similarity module which computes the predictions based on the updated representations mentioned above:
    \begin{align*}
        \Ba &= \mathrm{softmax}\left(\Bk \left(\Bm', \BX' \right)\right) \\
        \hat y &= \sum_{n=1}^{N} a_n \ y_n,
    \end{align*}
    
    where $\hat y$ is the prediction for the query molecule.
    For the computation of the attention values $\Ba$, the softmax 
    function is used. $\Bk$ is a similarity metric, such as
    the cosine similarity. 
    
    \paragraph{Hyperparameter search.}
    All hyperparameters were selected based on manual tuning
    on the validation set. We report the explored hyperparameter 
    space in Table \ref{tab:hyperparameterSpace_iterref}. Bold values 
    indicate the selected hyperparameters for the final model.
    
        \begin{table}
        \centering
        \begin{threeparttable}[b]
        \caption{Hyperparameter space considered for the IterRefLSTM model selection. The hyperparameters of the best configuration are marked bold. }
        \label{tab:hyperparameterSpace_iterref}
        
        \begin{tabular}{ll}
        \toprule Hyperparameter & Explored values \\
        \midrule
            Molecule encoder & \\
            ~~\textbullet~ Number of hidden layers & \textbf{0}, 1, 2, 4  \\
            ~~\textbullet~  Number of units per hidden layer & \textbf{1024}, 4096 \\
            ~~\textbullet~  Output dimension & \textbf{512}, 1024 \\
            ~~\textbullet~  Activation function & ReLU, \textbf{SELU}\\
            ~~\textbullet~  Input dropout & \textbf{0.1} \\
            ~~\textbullet~  Dropout & \textbf{0.5} \\
            IterRef embedding layer & \\
            ~~\textbullet~  L & 1, \textbf{3}\\
            Similarity module:&\\
            ~~\textbullet~  Metric & cosine similarity, \textbf{dot product}, MinMax similarity\\
            ~~\textbullet~  Similarity space dimension & 512, \textbf{1024}\\
            Layer-normalization & False, \textbf{True} \\
            ~~\textbullet~  Affine & \textbf{False}, True \\
            Training &\\
            ~~\textbullet~  Learning rate & 0.0001, \textbf{0.001}, 0.01\\
            ~~\textbullet~  Optimizer & \textbf{Adam}, AdamW\\
            ~~\textbullet~  Weight decay & \textbf{0}, 0.0001 \\
            ~~\textbullet~  Batch size & \textbf{2048}, 4096 \\
        \bottomrule    
        \end{tabular}
        \end{threeparttable}
    \end{table}
    
\subsubsection{MHNfs: details and hyperparameters}
\label{appsec:mhnfs}
    The MHNfs consists of a molecule encoder, the context module, the cross-attention-module, and the similarity module.
    The molecule encoder is a fully-connected Neural Network, consisting of one layer with 1024 units.
    For the context module, a Hopfield layer with 8 heads is used and also the cross-attention module include 8 heads. \textcolor{black}{We chose a concatenation of ECFPs and RDKit-based descriptors as the inputs for the MHNfs model. Notably, the RDKit-based descriptors were pre-processed in a way that instead of raw values quantils, which were computed by comparing a raw value with the distributation of all FS-Mol training molecules, were used. All descriptors were normalized based on the FS-Mol training data.}

    \paragraph{Hyperparameter search.} 
    All hyperparameters were selected based on manual tuning
    on the validation set. We report the explored hyperparameter 
    space in Table \ref{tab:hyperparameterSpace_mhnfs}. Bold values 
    indicate the selected hyperparameters for the final model.
    Early stopping points for the different reruns are chosen based on the $\Delta$AUC-PR metric on the validation set. For the five reruns the early-stopping points, that were automatically chosen 
    by their validation metrics, were the checkpoints at 
    epoch 94, 192, 253, 253 and 309.
   
    \begin{table}
        \centering
        \begin{threeparttable}[b]
        \caption{Hyperparameter space considered for the MHNfs model selection. The hyperparameters of the best configuration are marked bold. }
        \label{tab:hyperparameterSpace_mhnfs}
        
        \begin{tabular}{ll}
        \toprule Hyperparameter & Explored values \\
        \midrule
            Molecule encoder & \\
            ~~\textbullet~ Number of hidden layers & \textbf{0}, 1, 2, 4  \\
            ~~\textbullet~  Number of units per hidden layer & \textbf{1024}, 4096 \\
            ~~\textbullet~  Output dimension & \textbf{512}, 1024 \\
            ~~\textbullet~  Activation function & ReLU, \textbf{SELU}\\
            ~~\textbullet~  Input dropout & \textbf{0.1} \\
            ~~\textbullet~  Dropout & \textbf{0.5} \\
            Context module (hopfield layer) & \\
            ~~\textbullet~  Heads & \textbf{8}, 16\\
            ~~\textbullet~  Association space dimension & 512 [512;2048]\\
            ~~\textbullet~  Dropout & 0.1, \textbf{0.5}\\
            Cross-attention module (transformer mechanism) & \\
            ~~\textbullet~  Heads & 1, \textbf{8}, 10, 16, 32, 64\\
            ~~\textbullet~  Number units in the hidden feedforward layer & \textbf{567} [512; 4096] \\
            ~~\textbullet~  Association space dimension & 1088 [512;2048]\\
            ~~\textbullet~  Dropout & 0.1, \textbf{0.5}, 0.6, 0.7\\
            ~~\textbullet~  Number of layers: & \textbf{1}, 2, 3\\
            Similarity module:&\\
            ~~\textbullet~  Metric & cosine similarity, \textbf{dot product}, MinMax similarity\\
            ~~\textbullet~  Similarity space dimension & 512, \textbf{1024}\\
            ~~\textbullet~  $\tau$ & 32 [20;45]\\
            Layer-normalization & False, \textbf{True} \\
            ~~\textbullet~  Affine & \textbf{False}, True \\
            Training &\\
            ~~\textbullet~  Learning rate & \textbf{0.0001}, 0.001, 0.01\\
            ~~\textbullet~  Optimizer & \textbf{Adam}, AdamW\\
            ~~\textbullet~  Weight decay & \textbf{0}, 0.0001 \\
            ~~\textbullet~  Batch size & \textbf{4096} \\
            ~~\textbullet~ Warm-up phase (epochs) & \textbf{5} \\
            ~~\textbullet~ Constant learning rate phase (epochs) & 25, \textbf{35}\\
            ~~\textbullet~ Decay rate & 0.994\\
            ~~\textbullet~ \textcolor{black}{Max. number of epochs} & \textbf{350}\\
        \bottomrule    
        \end{tabular}
        \end{threeparttable}
    \end{table}
    
\begin{figure}
        \centering
        \includegraphics[width=1\textwidth]{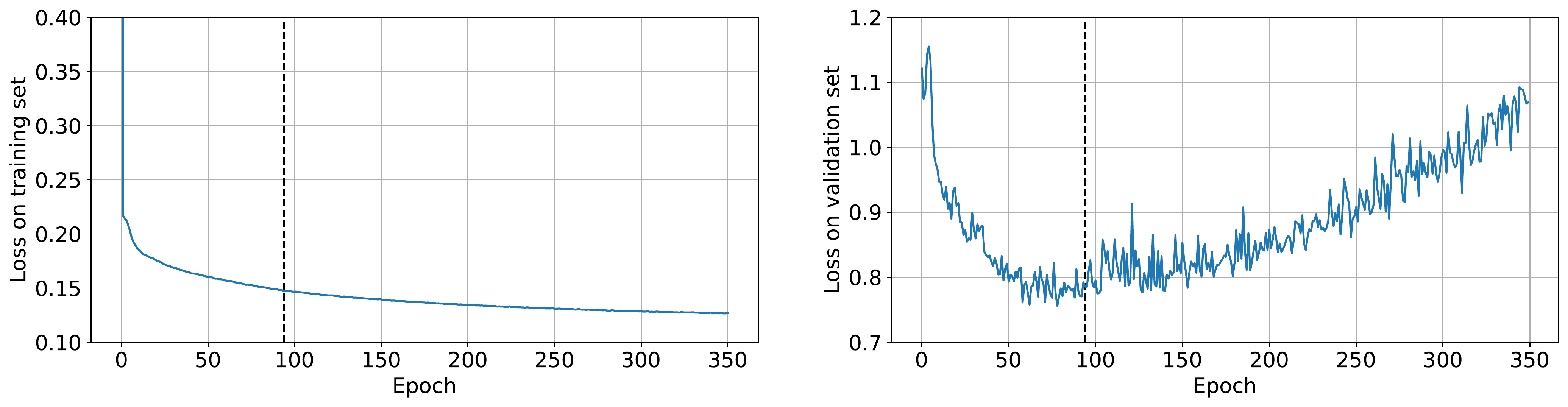}
        \caption{Exemplary MHNfs learning curve on FS-Mol. On the x-axis the number of epochs is displayed and on the y-axis the training loss (left) and the validation loss (right) is shown. The dashed line indicates the determined early-stopping point which is determined based on $\Delta$AUC-PR on the validation set.}
        \label{fig:learning_curve}
    \end{figure}

    \paragraph{\textcolor{black}{Model training}.}
    \textcolor{black}{Figure \ref{fig:learning_curve} shows the 
    learning curve of an exemplary training run of a MHNfs model on FS-Mol. 
    The left plot shows the loss on the training set and the right plot shows 
    the validation loss. The dashed line indicates the checkpoint of the 
    model which was saved and then used for inference on the test set, 
    whereas the stopping point was evaluated maximizing the $\Delta$AUC-PR 
    metric on the validation set.}
    
\begin{figure}
        \centering
        \includegraphics[width=1\textwidth]{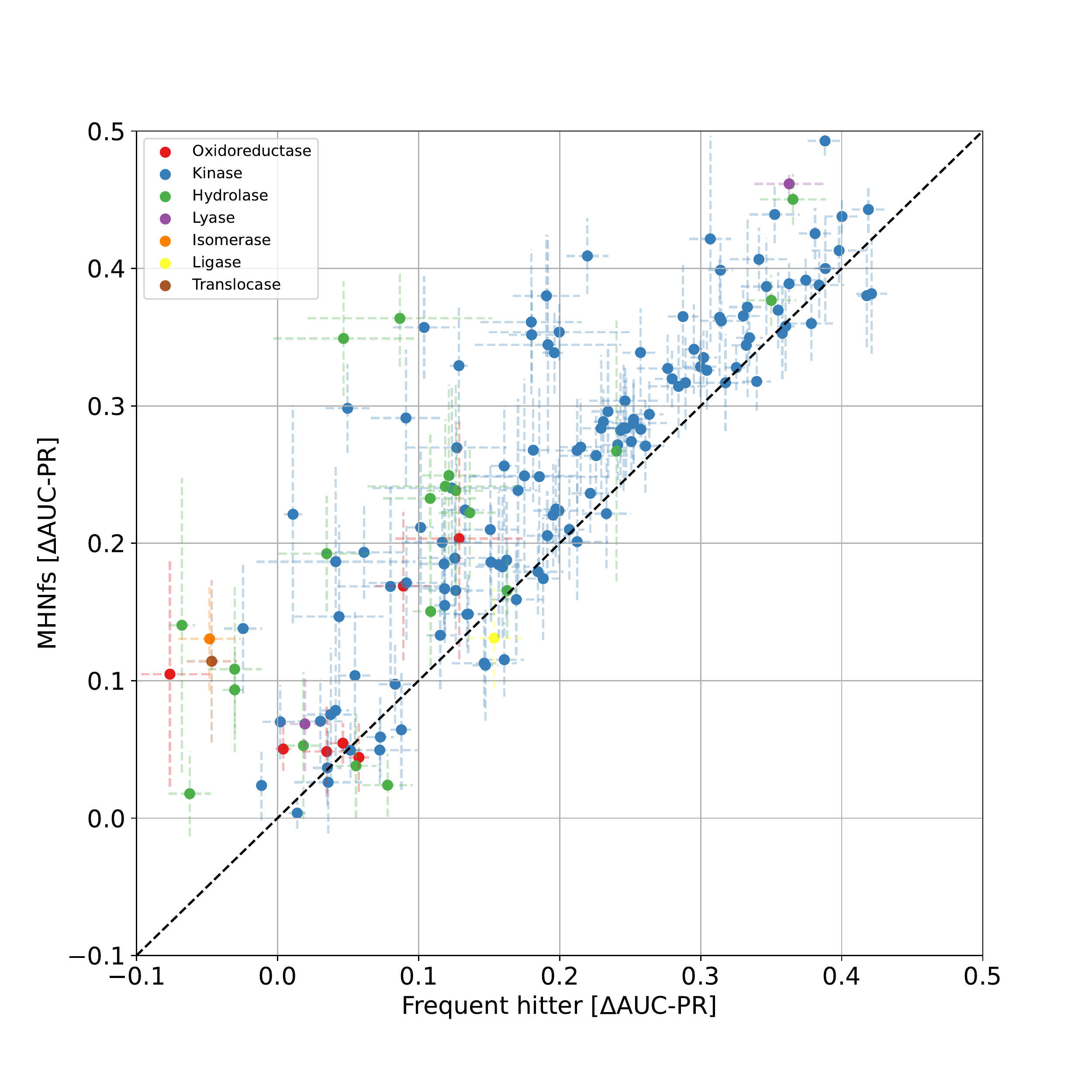}
        \caption{Performance comparison of MHNfs with the frequent hitter model. Each point refers to a task in the test set. Dashed lines indicate variablility across training reruns and different test support sets. The most points are located above the dashed line, which indicates that MHNfs performs better than den FH baseline at this task.}
        \label{fig:scatterplot_mhnfs_fh}
    \end{figure}

    \paragraph{\textcolor{black}{Performance improvements in comparison to a naive baseline.}}
    \textcolor{black}{Figure \ref{fig:scatterplot_mhnfs_fh} shows a task-wise performance comparison between MHNfs and the Frequent Hitter model. Each point indicates a task in the test set and is colored according to their super-class membership. In 132 cases the MHNfs outperforms the frequent hitter model. In 25 cases the frequent hitter model yields better performance.}
    
\subsubsection{\textcolor{black}{PAR: details and hyperparameters}}
\label{appsec:PAR}
\textcolor{black}{The PAR model \citep{wang2021property} includes a pre-trained GNN encoder, which creates initial embeddings for the query and support set molecules. These embeddings are fed into an attention mechanism module which also uses activity information of the support set molecules to create enriched representations. 
Another GNN learns relations between query and support set molecules.}

\paragraph{\textcolor{black}{Hyperparameter search}.}
\textcolor{black}{For details we refer to \citet{wang2021property} and \url{https://github.com/tata1661/PAR-NeurIPS21/blob/main/parser.py}. All hyperparameters were selected based on manual tuning on the validation set. The hyperparameter choice for Tox21 \citep{wang2021property} was used as a starting point. We report the explored hyperparameter space in Table \ref{tab:hyperparameterSpace_par}. Bold values indicate the selected hyperparameters for the final model. Notably, we just report hyperparameter choices which were different from standard choices. We used a training script provided by \citep{wang2021property}, which can be found here \url{https://github.com/tata1661/PAR-NeurIPS21}}.

    \begin{table}
        \centering
        \begin{threeparttable}[b]
        \caption{Hyperparameter space considered for the PAR model selection. The hyperparameters of the best configuration are marked bold.}
        \label{tab:hyperparameterSpace_par}
        
        \begin{tabular}{ll}
        \toprule Hyperparameter & Explored values \\
        \midrule
            Training & \\
            ~~\textbullet~ Meta learning rate & $1.0\cdot10^{-05}$, \boldsymbol{$1.0\cdot10^{-04}$}, $1.0\cdot10^{-03}$, $1.0\cdot10^{-02}$\\
            ~~\textbullet~ Inner learning rate & 0.01, \textbf{0.1}\\
            ~~\textbullet~ Update step & \textbf{1}, 2\\
            ~~\textbullet~ Update step test & \textbf{1}, 2\\
            ~~\textbullet~Weight decay & \boldsymbol{$5.0\cdot10^{-05}$}, $1.0\cdot10^{-03}$\\
            ~~\textbullet~Epochs & 200000\\
            ~~\textbullet~Eval. steps & 2000\\
            Encoder & \\
            ~~\textbullet~ Use pre-trained GNN & \textbf{yes}, no\\
            Attention-based module & \\
            ~~\textbullet~ Map dimension & 128, \textbf{512}\\
            ~~\textbullet~ Map layer & \textbf{2}, 3 \\
            ~~\textbullet~ Pre fc layer & \textbf{0}, 2\\
            ~~\textbullet~ Map dropout & \textbf{0.1}, 0.5\\
            ~~\textbullet~ Context layer & \textbf{2}, 3, 4\\
            Relation graph & \\
            ~~\textbullet~ Hidden dimension & 8, 128, \textbf{512}\\
            ~~\textbullet~ Number of layers & 2, \textbf{4}\\
            ~~\textbullet~ Number of layers for relation edge update & 2, \textbf{3} \\
            ~~\textbullet~ Batch norm & yes, \textbf{no} \\
            ~~\textbullet~ Relation dropout 1 & \textbf{0}, 0.25, 0.5 \\
            ~~\textbullet~ Relation dropout 2 & \textbf{0.2}, 0.25, 0.5 \\
        \bottomrule    
        \end{tabular}
        \end{threeparttable}
    \end{table}

\subsection{\textcolor{black}{Details on the FS-Mol benchmarking experiment}}
\label{appsec:fsmol}
\textcolor{black}{
This section provides additional information for the FS-Mol benchmarking experiment (see Section \ref{sec:experiments}).}

\textcolor{black}{
\textbf{Memory-based baselines.}
}
\textcolor{black}{The Classic Similarity Search can be considered as a method with associative memory, where the label is retrieved from the memory. Notably, for this method, the associative memory is very limited since it is the support set. Siamese Networks, analogously to the Classic Similarity Search, retrieve the label from a memory, whereby the similarities are determined in a learned space. Also, the IterRefLSTM-based method can be seen as having a memory, whereby the LSTM controls storing and removing information from the training data by the input and the forget gate. In NLP, kNN-type memories are currently used. Conceptually, they are very similar to the Modern Hopfield Networks, setting the number of heads to one and choosing a suitable value for $\beta$.
}

\begin{figure}
        \centering
        \includegraphics[width=1\textwidth]{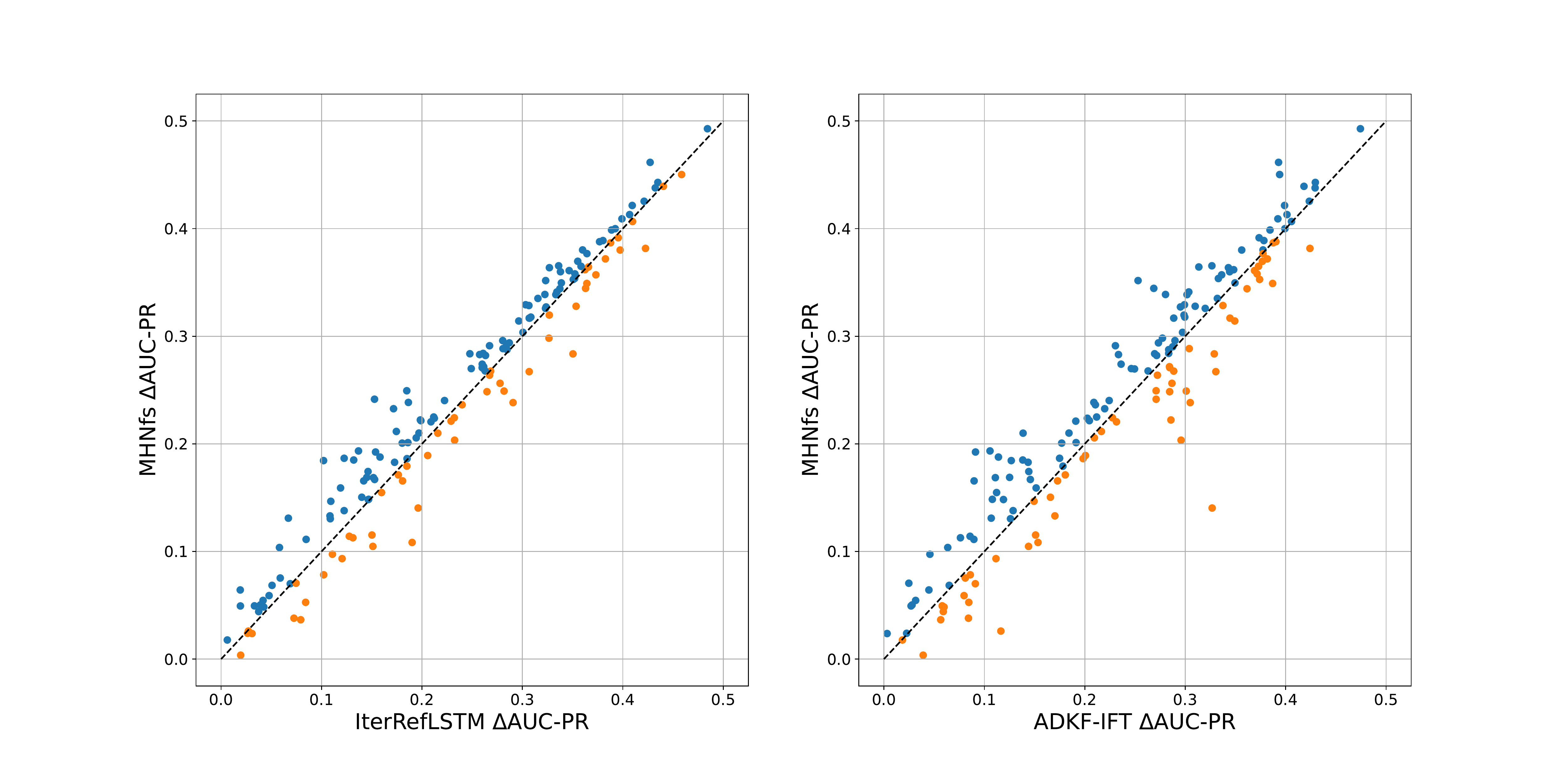}
        \caption{\textcolor{black}{Task-wise model comparison. The left scatterplot shows a comparison between MHNfs and the IterRefLSTM-based method and the right scatterplot shows a comparison between MHNfs and ADKF-IFT. Each dot refers to a task in the test set. For tasks on which the MHNfs performs better related dots are colored blue; otherwise the dots are colored orange.}}
        \label{fig:taskwise_comparison}
\end{figure}

\textcolor{black}{
\textbf{Results.}
}
\textcolor{black}{The reported performance metrics comprise three different sources of variation, namely variation across different tasks, variation across different support sets during test time, and variation across different training reruns. While error bars in Table \ref{tab:results_fsm} report variation across tasks, error bars in Table \ref{tab:fsmol_retrainerror} report variation across training reruns. For ADKF-IFT, the authors provided error bars for every single test task. Based on these error bars we sampled performance values to be able to compare ADKF-IFT with the MHNfs training reruns. Figure \ref{fig:taskwise_comparison} shows a task-wise model comparison between a) MHNfs and the IterRefLSTM-based method and b) MHNfs and ADKF-IFT. For a) MHNfs performs better on 106 of 157 tasks and therefore significantly outperforms the IterRefLSTM-based method (binomial test $p$-value $6.8\cdot 10^{-6}$). For b) MHNfs performs better on 98 tasks and therefore significantly outperforms ADKF-IFT (binomial test $p$-value $0.001$), too. Notably, ADKF-IFT performs better on non kinase-targets which can be seen in Table \ref{tab:results_fsm}.
}
\begin{table}
    \centering
    \begin{threeparttable}[b]
      \caption{\textcolor{black}{Results on FS-Mol [$\Delta$AUC-PR ]. The error bars represent standard deviation across training reruns.}
        \label{tab:fsmol_retrainerror}}
      \begin{tabular}{lcccc}
        \toprule
        Method & $\Delta \text{AUC-PR}$\\
        \midrule
        ADKF-IFT \citep{chen2022meta} &   .234 $\pm$ .001\\
        IterRefLSTM \citep{altae2017low} &   .234  $\pm$ .002\\
        MHNfs &   .241 $\pm$ .005\\
        \bottomrule
      \end{tabular}
    \end{threeparttable}
\end{table}

\subsection{Details on the ablation study}
\label{appsec:ablation}
The MHNfs has two new main elements compared to 
the most similar previous state-of-the art method IterRefLSTM, 
which are the context module and the cross-attention-module. 
In this ablation study we aim to investigate 
i) the importance of all design elements, which are the context module, 
the cross-attention module, and the similarity module, 
and ii) the superiority of the cross-attention module 
compared to the IterRefLSTM module.

\subsubsection{Ablation study A: comparison against IterRefLSTM}

\begin{table}
    \centering
    \begin{threeparttable}[b]
      \caption{Results of the ablation study on FS-Mol [AUC,
$\Delta$AUC-PR ]. The error bars represent standard deviation across training reruns and draws of support sets. The $p$-values indicate whether the difference between two models in consecutive rows is significant.}
        \label{tab:ablation_study_fsmol}
      \begin{tabular}{lcccc}
        \toprule
        Method & AUC & $\Delta \text{AUC-PR}$ & $p_{\mathrm{AUC}}$\tnote{a} & $p_{\Delta\mathrm{AUC-PR}}$\tnote{a} \\ \midrule
        MHNfs (CM+CAM+SM) &  .739 $\pm$ .005 & .241 $\pm$ .006 &  & \\
        MHNfs -CM &  .737 $\pm$ .004 & .240  $\pm$ .005 & 0.030 & 0.002\\
        MHNfs -CM -CAM  &  .719 $\pm$ .006 & .223 $\pm$ .006 & < 1.0e-8 & <1.0e-8\\
        Similarity Search &  .604 $\pm$ .003 & .113 $\pm$ .004 & <1.0e-8  & < 1.0e-8 \\
        \hline
        IterRefLSTM \citep{altae2017low}\tnote{b}&  .730 $\pm$ .005 & .234 $\pm$ .005 & <1.0e-8 & 8.73e-7\\
        \bottomrule
      \end{tabular}
      \begin{tablenotes}
        \item [a] paired Wilcoxon rank sum test
        \item [b] IterRefLSTM is compared to MHNfs -CM
      \end{tablenotes}
    \end{threeparttable}
\end{table}

For a fair comparison between the cross-attention module and 
the IterRefLSTM we used a pruned MHN version ("MHNfs -CM") 
which has no context module and compared it with the IterRefLSTM model.
The evaluation includes five training reruns each and ten different support set samplings.
The results, reported as the mean across training reruns and support sets, can be seen in Table \ref{tab:ablation_study_fsmol}.
We performed a paired Wilcoxon rank sum test for both the AUC and the $\Delta$AUC-PR metric.
Both $p$-values indicate high significance.

\subsubsection{Ablation study B: all design elements}
\label{appsec:domain_shift_alldesign}
We evaluate the performance of all main elements within the MHNfs, 
which are the context module, the cross-attention module,  
the similarity module and the molecule encoder.
For this analysis, we start with the complete MHNfs 
which includes all modules and report AUC and $\Delta$AUC-PR performance values.
Then, we iteratively omit the individual modules, 
measuring whether there is a significant performance 
difference with and without the module.
Table \ref{tab:ablation_study_fsmol} shows the results, 
where performance values for the full MHNfs, a MHNfs model 
without the context module ("MHNfs -CM") and a MHNfs module 
without the context and the cross-attenion module ("MHNfs -CM -CAM") is included.
Notably, the model without the context module and without 
the cross-attention module just consists of a 
learned molecule encoder and the similarity module.
We evaluted the impact of the learned molecule encoder 
by replacing it with a fixed encoder, which maps a 
molecule to its descriptors.
The model with the fixed encoder is a classic 
chemoinformatics method which is called Similarity Search \citep{cereto2015molecular}.

For the evaluation, we performed five training 
reruns for every model and sampled ten different 
support sets for every task.
Table \ref{tab:ablation_study_fsmol} shows the 
results in terms of AUC and $\Delta$AUC-PR.
We performed paired Wilcoxon rank sum tests on 
both metrics, comparing two methods in consecutive rows in the table.
The table shows that every module has a significant 
impact as omitting a module results in a significant 
performance drop.
The comparison between the MHNfs version without 
the context module and without the cross-attention 
module with the Similarity Search showed 
a significant superiority of the learned 
molecule encoder in comparison to the fixed encoder.

\subsubsection{Ablation study C: Under domain shift on Tox21}
\label{appsec:domain_shift_tox21}
Referring to Section \ref{appsec:domain_shift_alldesign}, 
the context module and the cross-attention module 
showed their importance for the global architecture.
This importance gets even more pronounced for the 
domain shift experiment on Tox21 as one can see 
in Table~\ref{tab:ablation_study_c}.

Again, five training reruns and ten support set 
draws are used for evaluation.
Including the context module  makes a clear 
and significant difference 
for both metrics AUC and $\Delta$AUC-PR.

\begin{table}
    \centering
    \begin{threeparttable}[b]
      \caption{Results of the ablation study on Tox21 [AUC,
$\Delta$AUC-PR ]. The error bars represent standard deviation across training reruns and draws of support sets. The $p$-values indicate whether a model is significantly different to the MHNfs in terms of the AUC and $\Delta$AUC-PR metric.}
      \label{tab:ablation_study_c}
      \begin{tabular}{lcccc}
        \toprule
        Method & AUC & $\Delta \text{AUC-PR}$ & $p_{\mathrm{AUC}}$\tnote{a} & $p_{\Delta\mathrm{AUC-PR}}$\tnote{a} \\ \midrule
        MHNfs (CM+CAM+SM) &  .679 $\pm$ .018 & .073 $\pm$ .008 &  & \\
        MHNfs -CM &  .662 $\pm$ .028 & .069  $\pm$ .012 & 6.28e-8 & 0.002\\
        MHNfs -CM -CAM &  .640 $\pm$ .018 & .057  $\pm$ .009 & <1.0e-8 & <1.0e-8\\
        Similarity Search &  .629 $\pm$ .015 & .061  $\pm$ .008 & <1.0e-8 & <1.0e-8\\
        IterRefLSTM &  .664 $\pm$ .018 & .067  $\pm$ .008  & 2.53e-6 & 3.38e-5\\
        \bottomrule
      \end{tabular}
      \begin{tablenotes}
        \item [a] paired Wilcoxon rank sum test
      \end{tablenotes}
    \end{threeparttable}
\end{table}

\subsection{\textcolor{black}{Details on the domain shift experiments}}
\textcolor{black}{
This section provides additional information for the Domain shift experminet on Tox21.
}

\textcolor{black}{
\textbf{Results.}
}
\label{appsec:domain-shift}
\textcolor{black}{The reported performance metrics comprise three different sources of variation, namely variation across different tasks, variation across different support sets during test time, and variation across different training reruns. While error bars in Table \ref{tab:results_tox21} report variation across both, drawn support sets and training reruns, error bars in Table \ref{tab:results_tox21_2} just report variation across training reruns. Notably, for the Similarity Search, the performance values do not vary since the model does not include any trainable parameters.
}

\begin{table}
    \centering
    \begin{threeparttable}[b]
      \caption{\textcolor{black}{Results of the domain shift experiment on the Tox21 dataset [AUC, $\Delta \text{AUC-PR}$]. The best method is marked bold. Error bars represent standard deviation across training reruns \label{tab:results_tox21_2}}}
      \begin{tabular}{lcccc}
        Method & AUC & $\Delta \text{AUC-PR}$  \\ \midrule
        Similarity Search (baseline)\tnote{a}  &  .629 $\pm$ .000 & .061 $\pm$ .000\\ 
        IterRefLSTM\hspace{0.5em}\citep{altae2017low} &  .664 $\pm$ .004  & .067 $\pm$ .001 \\
        MHNfs  &  \textbf{.679} $\pm$ .009 & \textbf{.073} $\pm$ .003 \\
        \bottomrule
      \end{tabular}
      \begin{tablenotes}
        \item [a] The Similarity Search does not include any learned parameters. Therefore, there is no variability across training reruns.
    \end{tablenotes}
    \end{threeparttable}
\end{table}

\subsection{Generalization to different support set sizes}
\label{appsec:sss}

\begin{figure}
        \centering
        \includegraphics[width=1\textwidth]{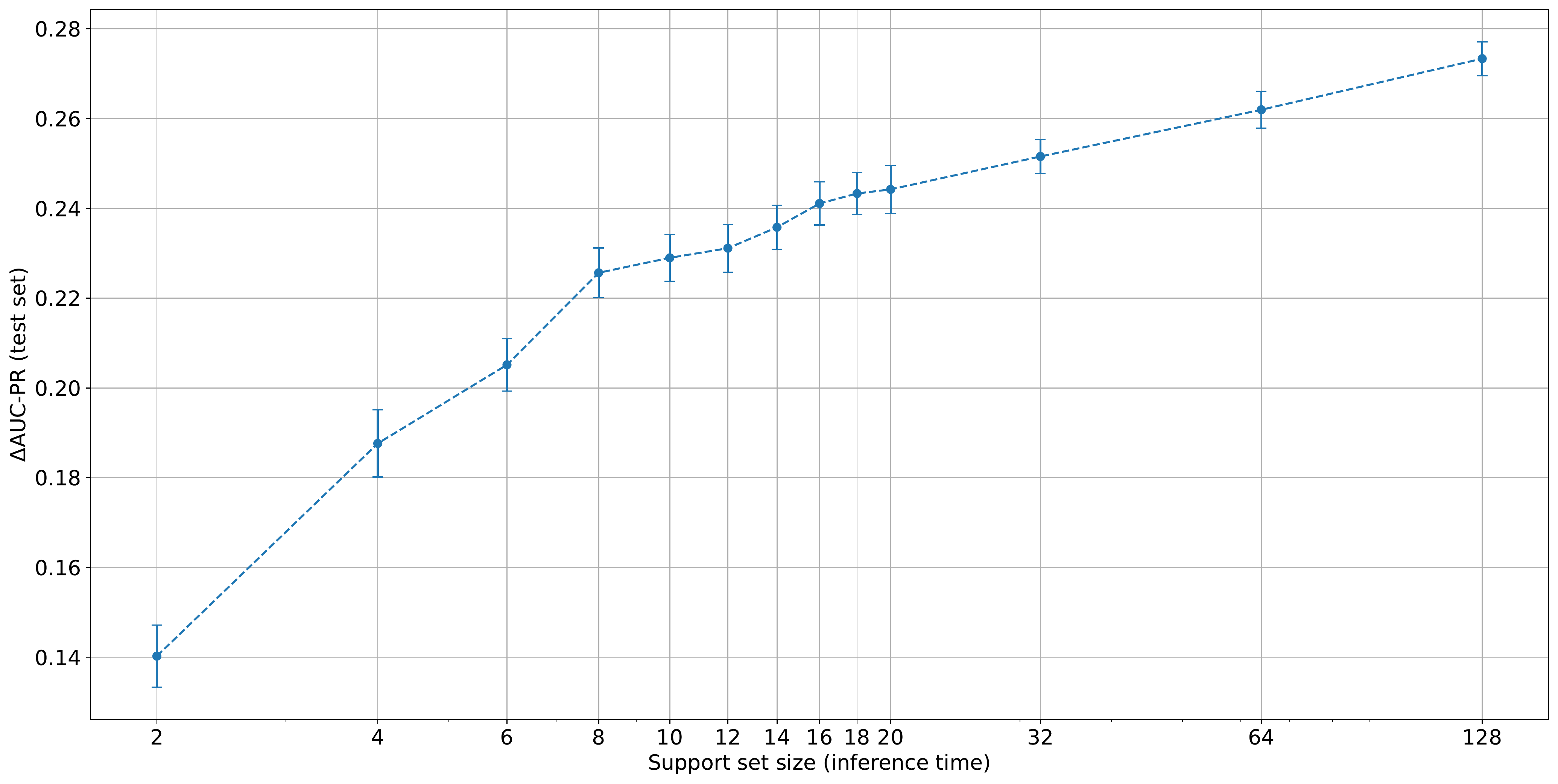}
        \caption{Performance of MHNfs for different support set sizes during inference time. The MHNfs models are trained with support sets of the size 16.}
        \label{fig:different_supportSetSizes}
\end{figure}

In the following section, we test the ability of MHNfs to generalize to different 
support set sizes. During training in the FS-Mol benchmarking setting, the 
MHNfs model has access to support sets of size 16. However, at 
inference, the support set size might be different. Figure~\ref{fig:different_supportSetSizes} provides 
performance estimates of the support-set-size-16 MHNfs models
on other support set sizes. 
Note that the estimates could be seen as approximate lower bounds of the 
predictive performance on settings 
with different support set sizes (y-axis labels). 
For a model used in production or in a real-world drug discovery 
setting, MHNfs should be trained with varying support set sizes
that resemble the distribution of real drug discovery projects.

\citet{triantafillou2019meta} analysed the performance of different few-shot models across different support set sizes.
Their analysis showed that in very-low-data settings embedding-based methods, namely Prototypical Networks and fo-Proto-MAML, performed best.
In contrast to that, finetuning-based method significantly profit from larger support set sizes \citep{triantafillou2019meta}.

MHNfs is an embedding-based method and performs -- in accordance with the findings mentioned above \citep{triantafillou2019meta} -- well for small support set sizes (see Table \ref{tab:results_fsm}). 
Following \citet{triantafillou2019meta}, it is exactly the settings related to these smaller support set sizes, e.g. a support set size of 16, which are suitable for MHNfs. For large support set sizes, e.g. 64 or 128, we point to the work done by \citet{chen2022meta}, in which the fine-tuning method ADKF-IFT achives an $\Delta$AUC-PR-score $>0.3$ for large support set sizes. 

\subsection{Generalization to different context sets}
In this section, we test the ability of MHNfs to generalize to different context sets. While the FS-Mol training split is used as a context during training, we assessed
whether our model is robust to different context sets for inference. To this end
we preprocessed the GEOM dataset \citep{axelrod2022geom}
from which we used 
100,000
molecules that passed all pre-processing checks. From this set, we sample 10,000
molecules as context set for MHNfs. Because GEOM contains drug-like molecules, 
similar to FS-Mol the predictive performance remains stable (see Table~ \ref{tab:context_performances}). 

\begin{table}
    \centering
    \begin{threeparttable}[b]
      \caption{MHNfs performance for different context sets [$\Delta$AUC-PR ]. The error bars represent standard deviation across training reruns and draws of support sets.}
      \label{tab:context_performances}
      \begin{tabular}{lc}
        \toprule
        Dataset used as a context  & $\Delta \text{AUC-PR}$\\ \midrule
        FS-Mol \citep{stanley2021fs} & .2414 $\pm$  .006\\
        GEOM \citep{axelrod2022geom} & .2415 $\pm$ .005\\
        \bottomrule
      \end{tabular}
    \end{threeparttable}
\end{table}

\subsection{\textcolor{black}{Details and insights on the context module}}
\label{appsec:cm_details}

\textcolor{black}{The context module replaces the initial representations of query and 
support set molecules by a retrieval from the context set. The context set is a large set 
of molecules and covers a large chemical space. The context module learns how to replace 
the initial molecule embeddings such that the context-enriched representations are put in 
relation to this large
chemical space and still contains all necessary information for the similarity-based 
prediction part. Figure \ref{fig:context} shows the effect of the context module
for the MHNfs model. Extreme initial embeddings, such as the purple embedding on the 
right, are pulled more into the known chemical space, represented by the context 
molecules. Notably, the replacement described above is a soft replacement, because also 
the initial embeddings contribute to the context-enriched representations due to 
skip-connections.}

\subsection{Reinforcing the covariance structure in the data using
modern Hopfield networks}
\label{appsec:covariance}
We follow the argumentation of \citep[Theorem A3]{furst2021cloob} that 
retrieval from an associative memory of a MHN reinforces 
the covariance structure. 

Let us assume that we have one molecule 
embedding from the query set $\Bm \in \mathbb{R}^d$ and one
molecule embedding from the support set $\Bx \in \mathbb{R}^d$ and
both have been enriched with the context module
with memory $\BC \in \mathbb{R}^{d \times M}$(ignoring linear mappings):
\begin{align}
    \Bm' &= \BC \ \mathrm{softmax}( \beta \BC^T \Bm) \\
    \Bx' &= \BC \ \mathrm{softmax}( \beta \BC^T \Bx)
\end{align}

Then the similarity of the retrieved representations 
as measured by the dot product can be expressed 
in terms of covariances:

\begin{figure}[H]
        \centering
        \includegraphics[width=0.7\textwidth]{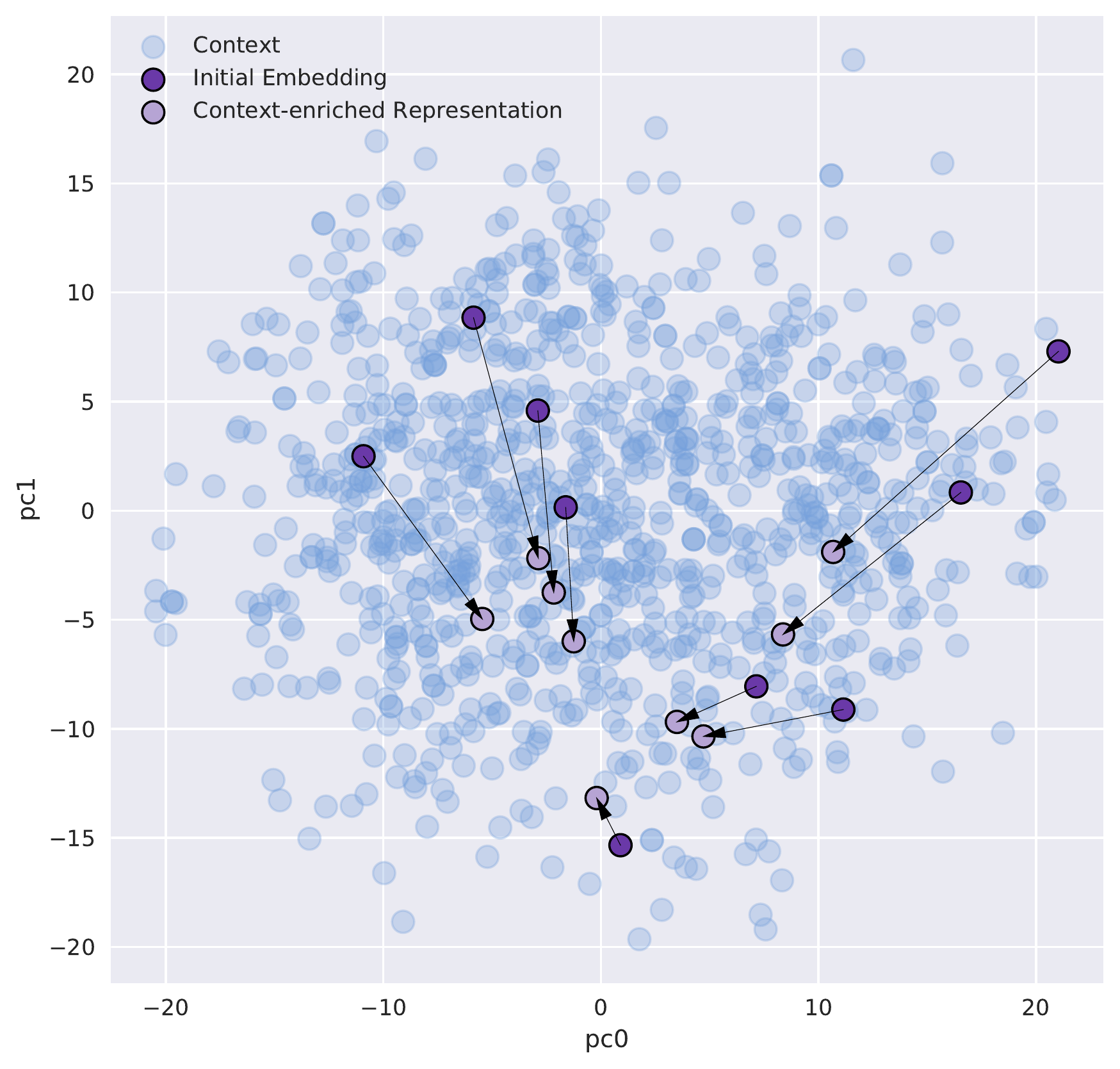}
        \caption{PCA-downprojection plot of molecule embeddings. 
        Each dot represents a molecule embedding, of which 
        the first two components are displayed on the x- and 
        y-axis. Blue dots represent context molecules. Dark purple dots
        represent 
        \textcolor{black}{initial embeddings for some exemplary molecules}, of which some exhibit extreme
        characteristics and are thus located away from the center.
        Arrows and light purple dots represent the enriched molecule 
        embeddings after the retrieval step. Especially molecules 
        from extreme positions are moved stronger to the center and 
        thus are more similar to known molecules after retrieval.}
        \label{fig:context}
\end{figure}

\begin{align}
    \Bm'^T \Bx' &=  \mathrm{softmax}( \beta \BC^T \Bm)^T \BC^T \BC \mathrm{softmax}( \beta \BC^T \Bx) = \\
    &= (\overline \Bc + \mathrm{Cov}(\BC, \Bm)^T \Bm)^T \ (\overline \Bc + \mathrm{Cov}(\BC, \Bx) \Bx),
\end{align}
where $\overline \Bc$ is the row mean of $\BC$ and following the 
\emph{weighted covariances} are used:
\begin{align}
    \mathrm{Cov}(\BC,\Bm)= \BC \mathrm{J^m} (\beta \BC \Bm) \BC^T  & \quad & \mathrm{Cov}(\BC,\Bx)= \BC \mathrm{J^m} (\beta \BC \Bx) \BC^T.   \
\end{align}
$\mathrm{J^m}: \mathbb{R}^M \mapsto \mathbb{R}^{M \times M}$ is a mean Jacobian function
of the softmax \citep[Eq.(A172)]{furst2021cloob}.

The Jacobian $\rJ$ of  $\Bp = \soft (\beta \Ba)$ is
$\rJ(\beta \Ba) =  \beta \ \left( \diag(\Bp) - \Bp \Bp^T  \right)$.
\begin{align}
  \Bb^T \rJ(\beta \Ba)  \ \Bb \ &= \
  \beta \ \Bb^T \left( \diag(\Bp) \ - \ \Bp \ \Bp^T \right) \ \Bb
  \ = \ \beta \ \left( \sum_i p_i \ b_i^2 \ - \ \left( \sum_i p_i \ b_i \right)^2 \right) \ ,
\end{align}
this is the second moment minus the mean squared, which 
is the variance. 
Therefore, $\Bb^T \rJ(\beta \Ba)  \Bb$ is $\beta$ times the covariance of $\Bb$ if component $i$ is drawn 
with probability $p_i$ of the multinomial 
distribution $\Bp$. 
In our case the component $i$ is context sample $\Bc_i$.
$\mathrm{J^m}$ is the average of $\rJ(\lambda \Ba)$ over
$\lambda=0$ to $\lambda=\beta$.

Note that we can express the enriched representations using these covariance functions:
\begin{align}
    \Bm' &= (\overline \Bc + \mathrm{Cov}(\BC, \Bm)^T \Bm)\\
    \Bx' &= (\overline \Bc + \mathrm{Cov}(\BC, \Bx)^T \Bx),
\end{align}
which connects retrieval from MHNs with 
reinforcing the covariance structure of the data.

\subsection{\textcolor{black}{Discussion, limitations and broader inpact}}
\label{appsec:discussion}
In a benchmarking experiment, the architecture was assessed 
for its ability to learn accurate predictive models 
from small sets of labelled molecules and in this setting
it outperformed all other methods. 
In a domain shift study, the robustness and transferability
of the learned models has been assessed and again \textbf{MHNfs}
exhibited the best performance.
The resulting predictive models often reach an AUC larger 
than $.70$, which means that
enrichment of active molecules is expected \citep{simm2018repurposed}
when the models are used for virtual screening. 
It has not escaped our notice that the specific context module 
we have proposed could immediately be used 
for few-shot learning tasks in computer vision, 
but might be hampered by computational constraints. 

Effectively using the information stored in the training data for new tasks is not only a key for our context-module but also for a lot of other few-shot strategies like pre-training or meta-learning. For pre-training and meta-learning based approaches, this information is stored in the model weights, while the context module directly has access to it via an external memory. We believe that accessing this information directly via an external memory is benefitial in this setting because a) pre-training for small molecule drug discovery is a promising approach, but still comes with its own challenges \citep{xia2022systematic} and b) a meta-learning approach, like MAML, needs labeled data while Modern Hopfield Networks operate on unlabeled data and therefore might be able to give access to more comprehensive information in the data including unlabeled data points.

\textbf{Limitations.}
In the FS-Mol benchmark experiment, the runner-up method ADKF-IFT \citep{chen2022meta} performed better on non kinase-tasks. We hypothesize that we could improve the MHNfs performance for non kinase tasks by upsampling the other task sub-groups. 
While the implementation of our 
method is currently limited to 
small, organic drug-like molecules as inputs,
our conceptual approach can also be used 
for macro-molecules such as RNA, DNA or proteins. 
The output domain of our method
comprises biological effects, 
such that the prediction must be understood in that domain. 
Our method demands higher computational costs and
memory footprint
as other embedding-based methods because
of the calculations necessary for the context module.
While we hypothesize that our approach
could also be successful for similar data in the materials science
domain, this has not been assessed. 
Our study is also constrained
by a limited amount of hyperparameter search for all methods. 
Deep learning methods usually 
have a large number of hyperparameters, such 
as hidden dimensions, number of layers, learning rates, 
of which we were only able to explore the most important ones.
The composition and choice of the context set is also under-explored
and might be improved by selecting reference molecules with 
an appropriate strategy.

\textbf{Broader impact.}
\emph{Impact on machine learning and related scientific fields.}
We envision that with (a) the increasing availability of 
drug discovery and \textcolor{black}{material science datasets}, 
(b) further improved biotechnologies, 
and (c) accounting for characteristics of individuals, 
the drug and materials discovery process will be made more efficient.
For machine learning and artificial intelligence, the novel way 
in which representations are enriched with context might strengthen 
the general research stream to include more context into 
deep learning systems. 
Our approach also shows that such a system is more robust 
against domain shifts, which could be a step towards Broad AI \citep{chollet2019measure,hochreiter2022toward}.
\emph{Impact on society.} If the approach proves useful, 
it could lead to a faster and more cost-efficient drug discovery process. 
Especially the COVID-19 pandemic has shown that it is crucial for 
humanity to speed up the drug discovery process to few years or even months. 
We hope that this work contributes to this effort and eventually
leads to safer drugs developed faster.
\emph{Consequences of failures of the method.} 
As common with methods in machine learning, 
potential danger lies in the possibility that users rely too much on 
our new approach and use it without reflecting on the outcomes.
Failures of the proposed method would lead to 
unsuccessful wet lab validation and
negative wet lab tests.
Since the proposed algorithm does not directly suggest treatment or therapy,
human beings are not directly at risk of being treated with a harmful therapy.
Wet lab and in-vitro testing would indicate wrong decisions by the system.
\emph{Leveraging of biases in the data and potential discrimination.}
As for almost all machine learning methods, confounding factors, lab or batch
effects, could be used for classification.
\textcolor{black}{This might} lead to biases in predictions or uneven predictive performance
across different drug targets or bioassays.

\end{document}